\documentclass[12pt]{article}
\usepackage[british]{babel}
\def\AnswerYes{y}
\def\draftVersion{n}               
\def\feynVersion{n}                
\usepackage[T1]{fontenc}

\usepackage{cite}                         
\usepackage[dvips]{color}                 
\usepackage[final]{graphicx}              
\graphicspath{{figures/}{./}}

\usepackage{multirow}                     
\usepackage{amssymb}
\usepackage{amsmath}
\usepackage{xspace}                       
\ifx\feynVersion\AnswerYes
   \usepackage{feynmp}                    
\fi

\ifx\draftVersion\AnswerYes
   \usepackage[scriptsize]{myshowkeys}    
   \newcommand{\comment}[1]{%
     {\scriptsize \sffamily \bfseries #1} }
   \newcommand{\margin}[1]{%
     \marginpar{\scriptsize \sffamily \bfseries #1} }
   \newcommand{\drafty}{\textbf{Draft version \today} \hfill}
\else
   \newcommand{\comment}[1]{}
   \newcommand{\margin}[1]{}
   \newcommand{\drafty}{}
\fi
\newcommand{\disc}{\discretionary{}{}{}}
\newcommand{\absatz}{\vspace{2ex}\noindent}

\newcommand{\PR}{\textnormal{Phys.\ Rev.\ }}

\newcommand{\PRL}{\PR\textnormal{Lett.\ }}

\newcommand{\hq}{\hspace{0.5em}}

\newcommand{\ii}{\mathrm{i}}
\newcommand{\dd}{\mathrm{d}}

\newcommand{\deintdim}[2]{\frac{\dd^{#1}\! #2}{(2\pi)^{#1}}\;}

\newcommand{\kv}{\vec{k}}

\newcommand{\mpi}{\ensuremath{m_\pi}}

\newcommand{\MeV}{\ensuremath{\mathrm{MeV}}}
\newcommand{\fm}{\ensuremath{\mathrm{fm}}}
\newcommand{\EFTNoPion}{EFT(${\pi\hskip-0.55em /}$)\xspace}

\newcommand{\NXLO}[1]{N\ensuremath{{}^{#1}}LO\xspace}

\newcommand{\fourHe}{\ensuremath{{}^4\mathrm{He}}\xspace}

\newcommand{\LambdaNoPion}{\ensuremath{\Lambda_{\pi\hskip-0.4em /}}}

\newcommand{\calA}{\mathcal{A}} 
 
\newcommand{\calL}{\mathcal{L}}

\newcommand{\mytitle}[1]{\begin{center}\LARGE{\textbf{#1}}\end{center}}
\newcommand{\myauthor}[1]{\textbf{#1}}
\newcommand{\myaddress}[1]{\textit{#1}}
\newcommand{\mypreprint}[1]{\begin{flushright}#1\end{flushright}}

\def\larr#1{\stackrel{\leftarrow}{#1}}
\def\rarr#1{\stackrel{\rightarrow}{#1}}

\newcommand{\One}{1}

\newcommand{\SDALINAC}{\textsc{S-Dalinac}\xspace}
\newcommand{\HIGS}{HI$\gamma$S\xspace}

\begin{document}
%

\begin{titlepage}
  \setcounter{page}{0} \vspace*{-2 true cm}\mypreprint{
    \drafty
    arXiv:0803.1307\\
    29th February 2008 \\
    Revised version 13th March 2008\\
    Final version 21st May 2008\\
    Accepted by Physical Review \textbf{C}
  }
  
  
  \mytitle{Pion-less Effective Field Theory on \\[1ex]
    Low-Energy Deuteron Electro-Disintegration}
  
 \vspace*{0.0cm}

\begin{center}
  \myauthor{Stefan Christlmeier$^{a}$}\\[1ex]
  \emph{and}\\[1ex]
  \myauthor{Harald W.\ Grie\3hammer$^{a,b,}$}\footnote{Corresponding author;
    Email: hgrie@gwu.edu; permanent address: b}\\[2ex]
  
  \vspace*{0.3cm}
  
  \myaddress{$^a$
    Institut f{\"u}r Theoretische Physik (T39), Physik-Department,\\
    Technische Universit{\"a}t M{\"u}nchen, D-85747 Garching, Germany}
  \\[2ex]
  \myaddress{$^b$ Center for Nuclear Studies, Department of Physics, \\The
    George Washington University, Washington DC 20052, USA}
  
  \vspace*{0.2cm}

\end{center}


\begin{abstract}
  In view of its relation to Big-Bang Nucleo-Synthesis and a reported
  discrepancy between nuclear models and data taken at \SDALINAC,
  electro-induced deuteron break-up ${}^2H(e,e^\prime p)n$ is studied at
  momentum transfer $q< 100$~MeV and close to threshold in the low-energy
  nuclear Effective Field Theory without dynamical pions, \EFTNoPion. The
  result at \NXLO{2} for electric dipole currents and at NLO for magnetic ones
  converges order-by-order better than quantitatively predicted and contains
  no free parameter. It is at this order determined by simple, well-known
  observables. Decomposing the triple differential cross-section into the
  longitudinal-plus-transverse ($L+T$), transverse-transverse ($TT$) and
  longitudinal-transverse interference ($LT$) terms, we find excellent
  agreement with a potential-model calculation by Arenh\"ovel et al using the
  Bonn potential. Theory and data also agree well on $\sigma_{L+T}$.  There is
  however no space on the theory-side for the discrepancy of up to $30\%$,
  $3$-$\sigma$ between theory and experiment in $\sigma_{LT}$. From
  universality of \EFTNoPion, we conclude that no theoretical approach with
  the correct deuteron asymptotic wave-function can explain the data.
  Un-determined short-distance contributions that could affect $\sigma_{LT}$
  enter only at high orders, i.e.~at the few-percent level. We notice some
  issues with the kinematics and normalisation of the data reported.
\end{abstract}
\vskip 0.5cm
\noindent
\begin{tabular}{rl}
  Suggested PACS numbers:& \begin{minipage}[t][\height][t]{10.7cm}
    13.40.-f, 21.30.-x, 21.45.Bc, 25.10.+s, 25.30.Fj, 26.35.+c, 27.10.+h
                    \end{minipage}
                    \\[3ex]
                    Suggested Keywords: &\begin{minipage}[t]{10.7cm}
                      Effective Field Theory, universality, 
                      deuteron electro-disintegration,
                      big-bang nucleo-synthesis. 
                    \end{minipage}
\end{tabular}

\vskip 1.0cm

\end{titlepage}

\setcounter{footnote}{0}

\newpage

%

\section{Introduction}
\setcounter{equation}{0}
\label{sec:introduction}

The deuteron is the simplest nucleus, playing the same fundamental r\^ole in
Nuclear Physics as the hydrogen atom in Atomic Physics. Its electromagnetic
properties have been studied in great detail both theoretically and
experimentally -- amongst others in photo-disintegration (see
e.g.~\cite{ArenSanz,GilmanGross} for reviews) and electro-disintegration (see
e.g.~\cite{FabianAren,ArenLeideTom04} and references therein).  The latter
process has the advantage to allow for an independent variation of energy and
momentum transfer.  Most experiments have been performed at higher energy
transfers and are in good agreement with potential-model calculations.
However, one experiment~\cite{Exp,Expweb} at the \SDALINAC accelerator of TU
Darmstadt (Germany) examined in 2002 the triple-differential cross-section for
$d(e,e^\prime p)n$ at low momentum transfer ($<60$~MeV/c) and close to the
breakup threshold, concentrating on the decomposition into the different
structure functions. For the longitudinal-transverse interference
cross-section $\sigma_{LT}$, a discrepancy of about 30\% or up to 3
standard-deviations was discovered relative to the prediction by Arenh\"ovel
et al.~\cite{ArenLeideTom04,Exp,Aren}. As the discrepancy is less pronounced
with increasing momentum and energy transfer, this does not necessarily
contradict other experiments at higher transfers which are in good agreement
with the same potential-model calculations, see e.g.~\cite{Tamae}.

The disagreement raises however a serious question: The same reaction at the
real-photon point, i.e.~photo-induced low-energy deuteron breakup and
recombination $np\leftrightarrow d\gamma$, is the first and very sensitive
step in Big-Bang Nucleo-Synthesis BBN. Indeed, both the higher-energy r\'egime
of BBN-relevant photo-dissociation and the \SDALINAC experiment on
electro-dissociation are most sensitive to the same electric dipole transition
amplitude $E1_V$. Due to the great difficulty of very-low-energy experiments,
the lack of data for this reaction in the BBN-relevant energy-region
$E\in[20;300]\;\mathrm{keV}$ makes Nuclear Theory at present the sole
provider of input for the BBN network codes~\cite{nol00,cyb04}. While this has
also triggered further experiments at \SDALINAC~\cite{ExpNew} and the
High-Intensity $\gamma$-ray Facility \HIGS at
\textsc{Tunl}~\cite{Tornow:2003ze,weller}, the accuracy claimed by theoretical
approaches~\cite{Rupaknpdg,Ando:2005cz,ArenSanz,Marcucci:2004sq} is not yet
matched. In view of this, the \SDALINAC findings are the more troubling: How
reliably does theory understand the simplest nucleus at low energies, if one
of the conceptually simplest non-trivial processes seems to disagree with
data? Does this reflect deficits in our understanding of the long-range $NN$
force, meson-exchange currents, ``off-shell effects'' and cutoff-dependence,
or even gauge-invariance? The analysis in Ref.~\cite{Exp} notes that the
potential-model approach ``without meson-exchange, isobar and relativistic
currents accounts for 99\% of the final result for $\sigma_{LT}$, leaving no
room for further improvement'' and continues: ``At present, there exists no
explanation of this surprising result in the framework of conventional nuclear
theory. It is an open question whether an alternative interpretation can be
offered by effective field theory [\dots]''. In this work, we answer this
question by concluding that \emph{no consistent theoretical approach} can
accommodate the data of Ref.~\cite{Exp}.

We employ pion-less Effective Field Theory \EFTNoPion in the variant in which
the effective range is re-summed into the two-nucleon
propagator~\cite{2stooges_quartet,3stooges_quartet,pbhg,chickenpaper,Rupak:2001ci,BeaneSavrearr,Ando}.
\EFTNoPion is well-tested in low-energy reactions with up to three nucleons,
see e.g.~\cite{bira_review,HadrtoNuc,Phillips,BedvKolck} for reviews. Even
properties of \fourHe~\cite{Platter:2004qn,Platter:2004zs,Hammer:2006ct} and
$^6$Li~\cite{Stetcu:2006ey} are now studied successfully. In particular,
$np\to d\gamma$ was studied due to its relevance for
BBN~\cite{ChenRupSavnpdg,ChenSavnpdg,Ando:2005cz}, culminating in a \NXLO{4}
calculation by Rupak~\cite{Rupaknpdg} with a theoretical accuracy of <1\%. We
here study the \SDALINAC data~\cite{Exp} within the same framework, noting
that the lowest energy data show the biggest discrepancy and lie well within the
range of applicability of \EFTNoPion.  At momenta below the pion-mass, probes
cannot resolve the long-range part of the $NN$-potential as nonlocal. Thus,
the most general Lagrangean compatible with the symmetries of QCD can be built
out of contact interactions between nucleons as the only dynamical effective
hadronic low-energy degrees of freedom.  \EFTNoPion is the unique low-energy
version of QCD and allows often (as in this case) for simple, closed-form
results. The amplitude is expanded in a small, dimension-less parameter
$Q=\frac{p_\text{typ}}{\LambdaNoPion}$: the ratio between a typical
low-momentum scale $p_\text{typ}$ of the process involved, and the breakdown
scale $\LambdaNoPion$, set by the mass $\mpi$ of the pion as the lightest
particle which is not included as a dynamical effective degree of freedom.  This
allows one to estimate the theoretical uncertainties which are
induced by neglecting higher-order terms in the momentum expansion of all
forces. With a typical low-momentum scale set by the inverse deuteron size,
one obtains $Q\approx\frac{1}{3}$. Looking at the order-by-order convergence
of all applications considered so far, one finds this estimate to be quite
conservative and the actual convergence pattern to be much better.
Dimensional regularisation is employed in the two-nucleon system to
renormalise high-momentum parts of loops, preserving all symmetries at each
step.  External currents and relativistic effects are included in a
systematic, manifestly gauge-invariant way. No ambiguities arise from
meson-exchange currents or off-shell effects.

\absatz The presentation is organised as follows: After recalling the
essentials of \EFTNoPion in Sect.~\ref{sec:framework}, the cross-section and
individual contributions are given for deuteron electro-disintegration up to
next-to-next-to-leading order (\NXLO{2}) for electric dipole transitions
$E1_V$ and up to NLO for magnetic is-vector dipole transitions $M1_V$.
Section~\ref{sec:data} contains the detailed comparison with data, focusing
on: kinematics in the experiment (Sect.~\ref{sec:datakin}); the full
cross-section (Sect.~\ref{sec:datatotal}) and its decomposition into the
structure functions (Sect.~\ref{sec:structure}); data normalisation
(Sect.~\ref{sec:discrepancies}); and possible higher-order effects
(Sect.~\ref{sec:higherorder}). After discussing an experiment at slightly
higher energies~\cite{Tamae} in Sect.~\ref{sec:higherE}, the Conclusions of
Sect.~\ref{sec:conclusions} are followed by an Appendix with some useful loop
integrals and the explicit form of the hadronic matrix elements.  Further
background and details are available in S.~Christlmeier's Diplom
thesis~\cite{diplom}.

\section{Framework and Calculation}
\setcounter{equation}{0}
\label{sec:framework}

\subsection{Kinematics and Cross-Section}
\label{sec:kinematics}

\begin{figure}[!htb]
  \begin{center}
  \includegraphics[width=0.45\linewidth]{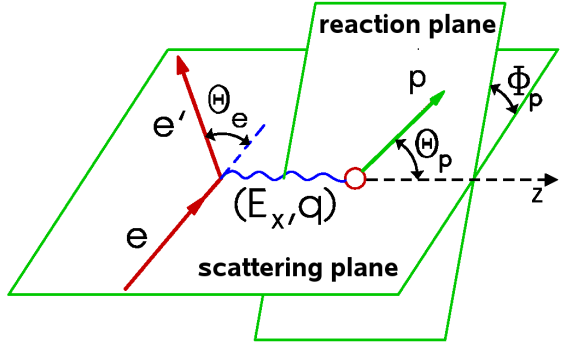}
  \caption{(Colour online) Kinematics of deuteron electro-disintegration,
    reproduced with kind permission of~\cite{Exp}. The electron kinematics
    refers to the lab frame, while the proton variables are defined in the
    centre-of-mass frame of the two-nucleon final state.}\label{ediskin}
  \end{center}
\end{figure}

The kinematic variables of the disintegration process $d(e,e^\prime p)n$ are
illustrated in Fig.~\ref{ediskin}.  Two frames of reference are conventionally
used, see e.g.~\cite{FabianAren,ArenLeideTom04,Arenhoevel:1992xu}.

On the one hand, the leptonic part $e\to\gamma^*e^\prime$ of the process is
conveniently discussed in terms of variables in the laboratory frame (the
deuteron rest frame), denoted by the superscript ``lab''. The four-momenta are
$(E_0^{\mathrm{lab}},\vec{k}^{\mathrm{lab}})$ for incoming, and
($E_e^{\mathrm{lab}},\vec{k}^{\prime\mathrm{lab}}$) for outgoing electrons.
The scattering angle between the momenta $\vec{k}^{\mathrm{lab}}$ and
$\vec{k}^{\prime\mathrm{lab}}$ of the incoming and outgoing electron is
$\Theta_e^{\mathrm{lab}}$. The energy and momentum transfer of the virtual
photon is
\begin{equation} 
  E_X^{\mathrm{lab}}:= \omega^{\mathrm{lab}}=
  E_0^{\mathrm{lab}}-E_e^{\mathrm{lab}}\;\;,\;\; \vec{q}^{\mathrm{lab}}=
  \vec{k}^{\mathrm{lab}}-\vec{k}^{\prime\mathrm{lab}}\;\;. \label{wlabqlab}
\end{equation}

The hadronic part of the process, $\gamma^*d \rightarrow pn$, is on the other
hand more conveniently calculated in its own rest frame, i.e.~in the
\emph{centre-of-mass frame of the outgoing nucleons} (and thus of the virtual
photon and deuteron). Variables in this frame carry no superscript. The
outgoing proton (neutron) has momentum $\vec{p}$ ($-\vec{p}$) and kinetic
energy $\frac{p^2}{2M}$, $p:=|\vec{p}|$.  The scattering angles in this system
are defined by $\vec{p}\cdot\vec{q}=pq\cos \Theta_p$ and
$\vec{p}\cdot(\vec{q}\times\vec{k})\propto\sin \Phi_p$. Thus, $\Theta_p$
is the angle between the virtual photon and the proton, and $\Phi_p$ is the
angle between the scattering plane, spanned by the incoming and outgoing
electron momenta, and the reaction plane, spanned by the proton and photon
momenta.

A boost along the momentum transfer $\vec{q}$ transforms between the two
frames:
\begin{equation}
  \beta = \frac{q^{\mathrm{lab}}}{M_d+\omega^{\mathrm{lab}}}\;\;,\;\;
  \gamma = \frac{1}{\sqrt{1-\beta^2}}\;\;,
\end{equation}
where $M_d=2M-B$ is the deuteron mass. Hence, the quantities defined in the lab
frame in (\ref{wlabqlab}) transform to the proton-neutron cm frame as:
\begin{equation}
  \omega =\gamma \omega^{\mathrm{lab}}-\beta \gamma
  q^{\mathrm{lab}}\;\;,\;\; q =\beta \gamma M_d\;\;.  
\end{equation}

We choose as the five independent variables of this process the energies
$E_0^{\mathrm{lab}}$, $E_e^{\mathrm{lab}}$ (or, equivalently,
$E_X^{\mathrm{lab}}=E_0^{\mathrm{lab}}-E_e^{\mathrm{lab}}$) and the scattering
angle $\Theta_e^{\mathrm{lab}}$ of the electrons in the lab frame, plus the
proton emission angles $\Theta_p$ and $\Phi_p$ in the proton-neutron cm frame.
The momentum of the outgoing proton is
\begin{equation} 
  p = \sqrt{(\omega-B)M+q^2 \frac{M}{2M_d}}\;\;. \label{protmom} 
\end{equation}
The momentum vectors are therefore parameterised such that $\vec{q}$ defines the $z$-direction,
\begin{eqnarray} 
  \vec{q}^{\mathrm{lab}} &= &(0,0,1) q^{\mathrm{lab}}\;\;,\\
  \vec{q} &= &(0,0,1) q\;\;,\\
  \vec{k}^{\mathrm{lab}} &= &(\sin \Theta_{0q}^{\mathrm{lab}},
  0, \cos \Theta_{0q}^{\mathrm{lab}}) k^{\mathrm{lab}}\;\;,\\
  \vec{k}^{\prime\mathrm{lab}} &= &(\sin \Theta_{eq}^{\mathrm{lab}}, 0, \cos
  \Theta_{eq}^{\mathrm{lab}}) {k}^{\prime\mathrm{lab}}\;\;,\\
  \vec{p} &= &(\sin \Theta_p \cos \Phi_p, \sin \Theta_p \sin \Phi_p, \cos
  \Theta_p) p\;\;, \label{eq:pvec}
\end{eqnarray} 
with the angles between the photon and the incoming or outgoing electrons
translating as
\begin{equation} 
  \cos \Theta_{0q}^{\mathrm{lab}} = \frac{k^{\mathrm{lab}}-
  {k}^{\prime\mathrm{lab}}\cos \Theta_e^{\mathrm{lab}}}{q^{\mathrm{lab}}}
  \;\;,\;\; \cos \Theta_{eq}^{\mathrm{lab}} = \frac{k^{\mathrm{lab}}\cos
  \Theta_e^{\mathrm{lab}}-{k}^{\prime\mathrm{lab}}}{q^{\mathrm{lab}}} \;\;.
\end{equation}
Since the transformation between these two frames is given by a boost along
$\vec{q}$, it does not affect the azimuthal angle
$\Phi_p\equiv\Phi_p^{\mathrm{lab}} $.

The electron momenta are $\kv_\mathrm{lab}^2 = E_{0\,\mathrm{lab}}^2-m_e^2$
and $\kv^{\prime2}_\mathrm{lab} = E_{e\,\mathrm{lab}}^2-m_e^2$. While the
electron mass $m_e=0.511$~MeV does not play a r\^ole for the experiment at
hand, we keep in mind applications to corners of phase-space where its effects
could be felt, e.g.~in back-scattering~\cite{ExpNew}.

The triple-differential cross-section is then derived from the amplitude
$\mathcal{A}$:
\begin{eqnarray}
  \frac{\dd^3\sigma}{\dd E_e^{\mathrm{lab}}\,\dd\Omega_e^{\mathrm{lab}}\,
    \dd\Omega_p} &=&
  \frac{{k}^{\prime\mathrm{lab}} p M_d M^2}{8 (2\pi)^5 (M+\frac{p^2}{2M}) 
  \sqrt{(M_d E_0^{\mathrm{lab}}+\vec{q}^{\mathrm{lab}}\cdot
    \vec{k}^{\mathrm{lab}})^2
  -M_d^2 m_e^2}}\;|\mathcal{A}|^2 
\label{cskin}\\
  &=&\frac{\dd^3}{\dd E_e^{\mathrm{lab}}\,\dd\Omega_e^{\mathrm{lab}}\,\dd\Omega_p} 
  (\sigma_L+\sigma_T+\sigma_{LT} \cos \Phi_p + \sigma_{TT} \cos 2\Phi_p)\;\;,
  \label{decomp}
\end{eqnarray}
where the spherical angles are
$\Omega_e^{\mathrm{lab}}=(\Theta_e^{\mathrm{lab}},
\Phi_e^{\mathrm{lab}}\equiv0)$ and $\Omega_p=(\Theta_p,\Phi_p)$. In the second
line, the dependence on the azimuthal proton emission angle $\Phi_p$ is
decomposed into the longitudinal-plus-transverse parts $\sigma_L+\sigma_T$ and
the interference terms $\sigma_{LT}$ and
$\sigma_{TT}$~\cite{FabianAren,ArenLeideTom04,Arenhoevel:1992xu}. For any
process to contribute to either of the interference terms, it must from
\eqref{eq:pvec} depend on those components of the outgoing proton momentum
$\vec{p}$ which are transverse to the photon momentum $\vec{q}$. This will
become important in the analysis later.

The scattering amplitude $\calA$ is rewritten in terms of the photon
propagator $D^{(\gamma)}_{\mu\nu}$, the hadronic current $J^\mu_\text{hadr}$ to be
calculated in \EFTNoPion below, and the leptonic current $l_\mu$ whose
contribution to lowest order in the fine-structure constant is easily found:
\begin{equation} 
  \mathcal{A} = l^\mu D^{(\gamma)}_{\mu\nu} J^\nu_\text{hadr}\;\;.
\end{equation}
Current conservation implies $
q_\mu l^\mu = 0 =q_\mu J^\mu_\text{hadr}$ and is at this order in the
fine-structure constant $\alpha$ equivalent to gauge invariance.

If the variables of Ref.~\cite{Exp} are interpreted as discussed in
Sect.~\ref{sec:datakin}, the \SDALINAC experiment was performed at two sets of
incident electron energies $E_0^{\mathrm{lab}}\in\{50;85\}\;\MeV$, a range of
photon energies $E_X^\text{lab}\in[8;16]\;\MeV$, electron scattering angle
$\Theta_e^{\mathrm{lab}}=40^\circ$ and azimuthal angle $\Phi_p=45^\circ$. This
leads to photon momentum transfers $q\in[32;65]\;\MeV$ and outgoing proton
momenta $p\in[74;106]\;\MeV$ in the proton-neutron cm-frame. These are the
relevant external low-momentum scales of the hadronic matrix element.

\subsection{Lagrangean and Parameter Fixing}
\label{sec:lagrangean}

\EFTNoPion allows one to address the questions raised by the \SDALINAC data in
a model-independent, systematic setting with a reliable estimate of the
theoretical uncertainties, free of ambiguities, like off-shell effects and cut-off
dependence.  As the parts of the effective Lagrangean of \EFTNoPion, its
power-counting rules and the parameter-fixing necessary for this work have
been discussed extensively by Beane and Savage~\cite{BeaneSavrearr}, we repeat
them here only briefly. The Feynman rules are also given in
\cite[App.~A]{diplom}.
\begin{eqnarray}
  \mathcal{L}_N &=&N^\dagger \bigg[\ii D_0+\frac{\vec{D}^2}{2M} 
  +\frac{e}{2M}(\kappa_0+\kappa_1\tau_3)\vec{\sigma}\cdot
  \vec{B}\bigg]N\;\;,
\label{LN} \\
  \mathcal{L}_s &= & -s_a^\dagger \bigg[\ii D_0 +\frac{\vec{D}^2}{4M}
  -\Delta_s\bigg]s_a 
  - y_s\left[s_a^\dagger N^T P_a^{(^1S_0)} N
    +\mbox{H.c.}\right]\;\;,  \label{Ls}\\
  \mathcal{L}_t &= &-t_i^\dagger \bigg[\ii D_0 +\frac{\vec{D}^2}{4M}
  -\Delta_t\bigg]t_i 
  - y_t\left[t_i^\dagger N^T P_i^{(^3S_1)} N
    +\mbox{H.c.}\right]\nonumber\\
  &  &-\frac{C_{sd}}{\sqrt{M\rho_d}}\!\!
  \left[\delta_{ix}\delta_{jy}-\frac{1}{3}\delta_{ij}\delta_{xy}\right]\!\!\!
  \left[t_i^\dagger
    (N^T\mathcal{O}_{xy,j}N)+\mbox{H.c.}\right] 
  -\frac{C_Q}{M\rho_d}t_i^\dagger
  \left[\ii D_0,\mathcal{O}_{ij}\right]t_j\;,  \label{Lsd} \\
  \mathcal{L}_{st}&=&\frac{eL_1}{M\sqrt{r_0\rho_d}} \left[t_i^\dagger s_3 B_i
    +\mbox{H.c.}\right]\;\;.  \label{Lst}
\end{eqnarray}
The one-nucleon Lagrangean $\mathcal{L}_N$ of the nucleon iso-doublet $N={p
  \choose n}$ with isospin-averaged mass $M=938.9$~MeV consists of two parts:
First, the kinetic term with minimal substitution, $D_{\mu} = \partial_\mu +
\ii eQA_\mu$, where $Q=\frac{1}{2}(\One+\tau_3)$ is the nucleon charge matrix and
$\alpha=\frac{e^2}{4\pi}=\frac{1}{137}$ the fine-structure constant. Second,
the interaction via the iso-scalar (iso-vector) magnetic moments,
$\kappa_0=0.44$ ($\kappa_1=2.35$) in nuclear magnetons. The spin (iso-spin)
Pauli matrices with vector (iso-vector) index $i=1,2,3$ ($a=1,2,3$) are
denoted by $\sigma_i$ ($\tau_a$).

The Lagrangean for the auxiliary di-baryon field $s_a$ ($t_i$) in the $^1S_0$
($^3S_1$) channel is $\mathcal{L}_s$
($\mathcal{L}_t$)~\cite{KaplanDibaryon,2stooges_quartet,3stooges_quartet,pbhg}.
As we probe only the $np$ system, $D_{\mu} =\partial_\mu + \ii eA_\mu$ for
both di-baryons. The projection operators are
$P_a^{(^1S_0)}=\frac{1}{\sqrt{8}}\sigma_2\tau_2\tau_a,\;
P_i^{(^3S_1)}=\frac{1}{\sqrt{8}}\sigma_2\sigma_i\tau_2$.  The auxiliary-field
parameters are matched to the effective-range expansion of $NN$ scattering
such that the effective range is re-summed to all powers. This speeds up
convergence and simplifies calculations of some higher-order
effects~\cite{2stooges_quartet,3stooges_quartet,pbhg,chickenpaper,Rupak:2001ci,BeaneSavrearr,Ando}:
\begin{equation}
  y_s = \frac{\sqrt{8\pi}}{M\sqrt{r_0}}\;,\;
  \Delta_s = \frac{2}{Mr_0}\left(\frac{1}{a_0}-\mu\right)\;,\;
  y_t = \frac{\sqrt{8\pi}}{M\sqrt{\rho_d}}\;,\;
  \Delta_t = \frac{2}{M\rho_d}\left(\gamma-\frac{\rho_d}{2}\gamma^2-\mu\right)
\end{equation}
The parameter $\mu$ encodes the linear cut-off divergence of individual
diagrams, calculated in the Power Divergence Subtraction scheme (PDS) version
of dimensional regularisation~\cite{KapSavWisePDS,KapSavWisePC}. It is
imperative for self-consistency that physical amplitudes are independent of
$\mu$. This is indeed the case.

For the spin-singlet, the scattering length and effective range of the $np$
system are probed in deuteron electro-disintegration: $a_0=-23.71\mbox{~fm},\;
r_0=2.73\mbox{~fm}$.  Parameters for the spin-triplet are determined from the
effective-range expansion about the observed real bound state, namely the
deuteron with binding energy $B=2.225\mbox{~MeV}$ (corresponding to a momentum
$\gamma=\sqrt{M B}=45.70\mbox{~MeV}$) and effective range
$\rho_d=1.764\mbox{~fm}$.  Therefore, the deuteron wave function shows the
correct exponential decay and normalisation $Z$ already at LO in
\EFTNoPion~\cite{PhilRupSav,BeaneSavrearr}:
\begin{equation} 
  \Psi_\text{deuteron}(r\rightarrow
  \infty) = \sqrt{\frac{Z}{2\pi\rho_d}} \frac{e^{-\gamma r}}{r}\;\;\mbox{ with }
  Z=\frac{\gamma \rho_d}{1-\gamma  \rho_d}
\label{deutwavefu} 
\end{equation}

The last two terms of $\mathcal{L}_t$ parameterise $SD$-wave mixing in the
spin-triplet, with
\begin{eqnarray}
  \mathcal{O}_{xy,j} 
  &\hspace{-2mm}= &\hspace{-2mm}-\frac{1}{4}
  \left(\larr{{D}}_x\larr{{D}}_yP_j^{(^3S_1)}
    +P_j^{(^3S_1)}\rarr{{D}}_x\rarr{{D}}_y
    -\larr{{D}}_xP_j^{(^3S_1)}\rarr{{D}}_y
    -\larr{{D}}_yP_j^{(^3S_1)}\rarr{{D}}_x\right) \;\;,\label{Oxyj}\\
  \mathcal{O}_{ij} &\hspace{-2mm}= 
  &\hspace{-2mm}-\left(\vec{D}_i\vec{D}_j-\frac{1}{3}\delta_{ij}\vec{D}^2\right)
\end{eqnarray}
operators which ensure that $\calL_{st}$ is manifestly gauge-invariant.  Their
strengths are determined from the asymptotic ratio $\eta_{sd}=0.0254$ of $D$-
and $S$-wave components of the deuteron wave function, and of the deuteron
quadrupole moment $\mu_Q=0.2859$~fm$^2$~\cite{BeaneSavrearr}:
\begin{equation}
  C_{sd} = \frac{6\sqrt{\pi}\eta_{sd}}{\sqrt{M} \gamma^2}\;\;,\;\;
  \mu_Q=2 Z
  \left[y_t\frac{C_{sd}}{\sqrt{M\rho_d}}\frac{M^2}{32\pi}\frac{2}{3\gamma}
    +\frac{C_Q}{M \rho_d}\right]  \label{quadrmom}
\end{equation}
Here, $C_{sd}$ contributes at LO, and $C_Q$ provides a NLO correction of about
$50\%$.

Finally, $\mathcal{L}_{st}$ in \eqref{Lst} parameterises transitions between
the $^1S_0$ and $^3S_1$ channels by a magnetic field acting on a di-baryon.
Its strength is determined by the thermal cross-section
$\sigma(E=\frac{p^2}{M}=1.264\cdot10^{-8}\mbox{~MeV}) = (334.2\pm0.5)$~mb
\cite{Cox} for radiative capture of neutrons on protons $np\to d\gamma$. At
thermal energies, it is dominated by iso-vectorial $M1_V$ transitions.
Magnetic-moment interactions are LO, and the parameter $L_1$ in
$\mathcal{L}_{st}$ enters as a NLO correction of about $50\%$, while electric
transitions are irrelevant.  The free parameter is thus determined from the
thermal cross-section~\cite{BeaneSavrearr}
\begin{equation}
  \sigma^{(M1_V)}= \frac{2\alpha Z(p^2+\gamma^2)^3}{pM^3} \;\left|
    \frac{1}{-\frac{1}{a}+\frac{1}{2}r_0p^2-\ii p} \right|^2  
  \left[\kappa_1\frac{\gamma-\frac{1}{a}+\frac{1}{2}r_0p^2}{p^2+\gamma^2} 
    +\frac{L_1}{2} \right]^2 \label{XM1V}
\end{equation}
as $L_1=-4.41\;\fm$~\cite{diplom} or $L_1=-4.0\;\fm$~\cite{BeaneSavrearr},
depending whether terms quadratic in $L_1$ are kept.
Both variants differ by $10\%\lesssim Q$, as expected from $L_1$ being a
higher-order correction.

\absatz We now state the power-counting rules of \EFTNoPion in the version in
which the effective-range parameters are treated as unnaturally large and thus
are kept at LO together with the scattering lengths~\cite{BeaneSavrearr}. The
typical low-energy centre-of-mass momentum $\vec{p}$ and kinetic energy $E/2$
of a nucleon is counted as
\begin{equation}
  |\vec{p}| \sim \sqrt{ME} \sim \gamma \sim \frac{1}{\rho_d} \sim
  \frac{1}{r_0} \sim \frac{1}{a_0} \sim \mu \sim Q\LambdaNoPion\;\;.  
  \label{PCbasic} 
\end{equation} 
Hence, the wave function renormalisation scales as $Z \sim 1$.  $SD$ mixing is
suppressed by $Q^2$ with respect to pure $S$-wave amplitudes because of the
two derivatives in \eqref{Lsd}. Finally, the ratio of the iso-scalar and
iso-vector magnetic moments is $\kappa_0/\kappa_1\approx 1/5\lesssim Q$ .
Therefore, $\kappa_0$ is neglected as numerically higher order in the
following~\cite{compton}.

Since the gauge field is minimally coupled, these rules are simple to extend:
The zero-component of the gauge field $A_\mu$ scales like an energy, $A^0 \sim
Q^2$; its 3-vector components like a momentum, $\vec{A} \sim Q$. As seen in
the previous Sub-Section, photon energies for the \SDALINAC kinematics of
deuteron electro-disintegration lie in the range
$E_X^\text{lab}\in[8;16]\;\MeV$, corresponding to $E_X\in[6;14]\;\MeV\sim
[Q^2\,\LambdaNoPion^2/M;\,Q^2\,\LambdaNoPion]$, photon momenta in the range
\mbox{$q\in[32;65]\;\MeV\sim Q\LambdaNoPion$}, and outgoing proton momenta in
the range \mbox{$p\in[74;106]\;\MeV\sim Q\LambdaNoPion$}, approaching
$\LambdaNoPion\approx\mpi$. Convergence must therefore be checked with care.

\subsection{Electric Contributions to \NXLO{2}}
\label{sec:el}
Electric dipole transition amplitudes $E1_V$ dominate not only the
photo-dissociation cross-section at the higher end of the energy range
relevant for BBN, but also electro-disintegration for $p \gtrsim 20$~MeV. To
\NXLO{2}, the hadronic currents have the structure
\begin{equation}
  J^{(E1_V)\,\mu}_{\text{hadr}} 
  = \ii e\sqrt{Z}\frac{1}{\sqrt{8}} (N_p^\dagger \sigma^i \sigma_2 N_n^*)
  \epsilon_{(d)}^j J_{ij}^\mu\;\;, \label{Ecurrentstruc} 
\end{equation} 
where the proton and neutron spinors $N_{p/n}$ are specified explicitly. The
indices on the current $J_{ij}^\mu$ indicate the polarisation $j$ of the
initial deuteron and the spin $i$ of the $pn$-triplet final state. Averaging
over spins and polarisations, the electric contribution to (\ref{cskin}) is
\begin{equation}
  |\mathcal{A}_{E1_V}|^2 =
  \frac{(4\pi\alpha)^2}{(\omega^2-q^2)^2}\frac{Z}{6}
  \left[\left({k^{\prime\mu}_\mathrm{lab}}J_\mu^{\dagger ij}\;
      {k^{\nu}_ \mathrm{lab}} J_\nu^{ij} +  (\mu\leftrightarrow\nu)\right) 
      - (k_\mu^\mathrm{lab} {k^{\prime\mu}_\mathrm{lab} }-m_e^2) 
      J^{\dagger ij}_\nu J^{\nu ij}\right]\;\;.
\end{equation}

\begin{figure}[!htb]
  \begin{center}
    \includegraphics*[scale=0.8]{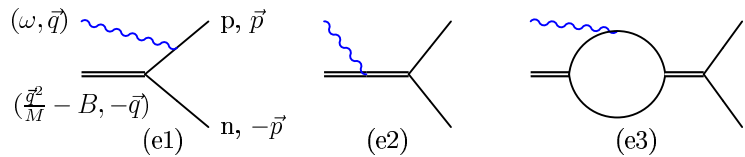}
    \caption{(Colour online) The LO electric contributions to the hadronic
      current. The double line denotes the spin-triplet di-baryon intermediate
      state. There is no NLO contribution.}\label{disLO}
  \end{center}
\end{figure}

The three diagrams contributing at leading order, $Q^0$, are shown in
Fig.~\ref{disLO}.
In the last two graphs, the photon couples to the deuteron which is at this
order a pure $\mathrm{S}$-wave state. These diagrams are therefore independent
of the direction of the outgoing proton momentum $\vec{p}$, and hence via
\eqref{eq:pvec} of the proton emission angle $\Phi_p$. They do thus not
contribute to the longitudinal-transverse and transverse-transverse parts of
the cross-section \eqref{decomp}. In contradistinction, the final-state
interaction in diagram (e1) depends on $\vec{p}$ and thus contributes to
$\sigma_{LT}$ and $\sigma_{TT}$.

\begin{figure}[!htb]
  \begin{center}
    \includegraphics*[scale=0.8]{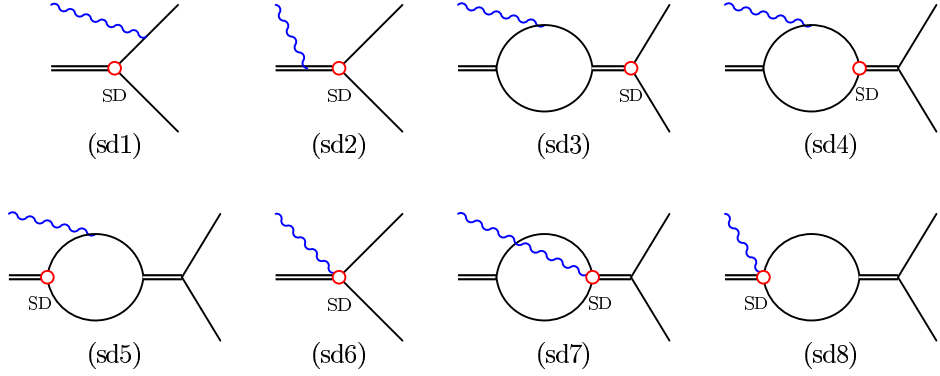}
    \caption{(Colour online) The order-$Q^2$ (\NXLO{2}) electric
      contributions. The $\text{SD}$-mixing vertex proportional to
      $C_\text{SD}$ is denoted by a circle; further notation as in
      Fig.~\ref{disLO}.}\label{sddiag}
  \end{center}
\end{figure}

There are no electric contributions at NLO. The first non-zero corrections,
listed in Fig.~\ref{sddiag}, appear at \NXLO{2}. All arise from
$\mathrm{SD}$-mixing proportional to $C_{SD}$ in the Lagrangean \eqref{Lst}.
Relative to the LO contributions, these diagrams are suppressed by two more
derivatives, i.e.~two more powers of $Q\approx \frac{1}{3}$ and thus should
contribute at most $\left(\frac{1}{3}\right)^2 \approx 10$\% of the LO terms.
The error one makes by truncating the series at this order should be
$\left(\frac{1}{3}\right)^3 \approx 4$\%. The convergence is indeed much
better because the asymptotic SD-ration is numerically
$\eta_{sd}=0.0254\approx Q^3$, i.e.~\NXLO{3}. Even at the \SDALINAC-kinematics
of photon momenta as large as $78$~MeV ($E_0^\mathrm{lab}=50$~MeV,
$E_X^\mathrm{lab}=9$~MeV, $\Theta_e^\mathrm{lab}=40^\circ$,
$\Phi_p=45^\circ$), they amount to not more than 1\% of the total; see the
discussion and figures in Sects.~\ref{sec:datatotal} and \ref{sec:structure}
as well as Ref.~\cite{diplom} for details.  Only the diagrams (sd1-3) and
(sd6) contribute to the interference terms $\sigma_{LT}$ and $\sigma_{TT}$ via
their final-state interaction or via the $D$-wave component induced into the
deuteron and $np$ final-state wave-function.

The resulting electric currents up to \NXLO{2} are detailed in
App.~\ref{app:electric}. At each order, they are manifestly gauge-invariant
and cutoff-independent.

\subsection{Magnetic Contributions to NLO}
\label{sec:mag}

\begin{figure}[!htb]
  \begin{center}
    \includegraphics*[scale=0.8]{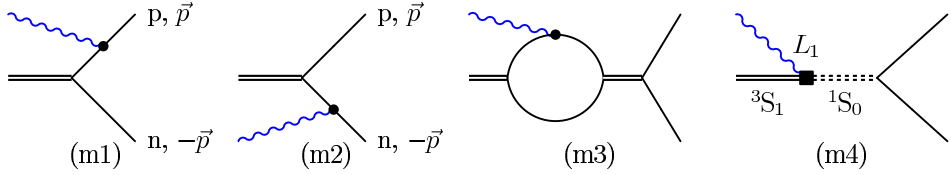}
    \caption{(Colour online) The magnetic contributions to the hadronic
      current at LO (m1-m3) and NLO (m4). The blob denotes a magnetic photon
      coupling via the iso-vectorial magnetic moment; the square the
      dibaryon-photon interaction \eqref{Lst} proportional to $L_1$; the
      dashed double-line the spin-singlet di-baryon intermediate state;
      further notation as in Fig.~\ref{disLO}.}\label{Mdiag}
  \end{center}
\end{figure}

Magnetic dipole transitions $M1_V$ dominate the inverse reaction $np\to
d\gamma$ at the lower end of the energy range relevant for BBN, but are
usually small at \SDALINAC kinematics. We therefore include them only to NLO,
Fig.~\ref{Mdiag}. The LO part is set by the iso-vector magnetic moment
$\kappa_1$; the only NLO graph, (m4), comes from the singlet-triplet coupling
\eqref{Lst}.

The hadronic currents from magnetic contributions can to this order be
written as
\begin{equation}
  J^{(M1_V)\,k}_{\text{hadr}} = e \sqrt{Z}
  \epsilon^{ijk}\epsilon_{(d)}^i\frac{1}{\sqrt{8}} 
  (N_p^\dagger\sigma_2N_n^*)\; J^j\;\;, \label{curstrucM}
\end{equation}
and the squared amplitude is after averaging over initial states:
\begin{eqnarray}
  |\mathcal{A}_{M1_V}|^2 &\hspace{-2mm}=
  &\hspace{-2mm}\frac{(4\pi\alpha)^2}{(\omega^2-q^2)^2}\frac{Z}{6} 
  \left(\delta^{jm}\delta^{kn}-\delta^{km}\delta^{jn}\right) \\
  &&\hspace{-4mm}\times \left[k^{\prime\,\mathrm{lab}}_n J^{\dagger j} 
    k_k^\mathrm{lab} J^m 
    +k^{\prime\,\mathrm{lab}}_k J^m k_n^\mathrm{lab}J^{\dagger j} 
    +(k_\mu^\mathrm{lab} {k^{\prime\mathrm{lab} \mu}}-m_e^2) 
    J^{\dagger j}J^m \delta^{kn} \right]\nonumber
\end{eqnarray}
The vertex where a magnetic photon couples to a di-baryon reduces the
amplitude largely, although it is formally NLO. Compared to the electric
amplitude, the magnetic one is suppressed by two orders of magnitude in the
energy r\'egime of the \SDALINAC experiment, except for the vicinity of
$\Theta_p=70^\circ$; see Figs.~\ref{cmcomp} to \ref{sigmadecomp2}.

It is not too difficult to see that the only angular dependence in the
magnetic amplitude comes from the $\vec{p}\cdot\vec{q}$ -terms in the
intermediate nucleon propagators, as the deuteron and outgoing $np$-system are
again pure $\mathrm{S}$-states in these diagrams. Thus, the amplitudes only
test the longitudinal part of $\vec{p}$. There is no $\Phi_p$-dependence, and
therefore magnetic transitions do at this order not contribute to
$\sigma_{LT}$ and $\sigma_{TT}$ interference terms \eqref{decomp}.

The explicit form of the resulting magnetic currents can be found in
App.~\ref{app:magnetic}. At each order, they are manifestly gauge-invariant
and cutoff-independent.

\section{Theory Confronts Data}
\setcounter{equation}{0}
\label{sec:data}

We now compare the results of \EFTNoPion to the \SDALINAC data reported
in~\cite{Exp,Expweb} and to the potential model calculation by H.~Arenh\"ovel
et al.~which is based on the Bonn-C $r$-space potential and includes
final-state effects, meson exchange currents, isobar configurations and (in
our case negligible) relativistic effects~\cite{ArenLeideTom04,Exp,Aren}.
Before we consider the energy- and angular dependence of the triple
differential cross-section and its decomposition, we first have to address a
subtle kinematical point.

\subsection{Kinematics, Again}
\label{sec:datakin}

\begin{figure}[!phtb]
  \begin{center}
  \includegraphics[width=0.5\linewidth]{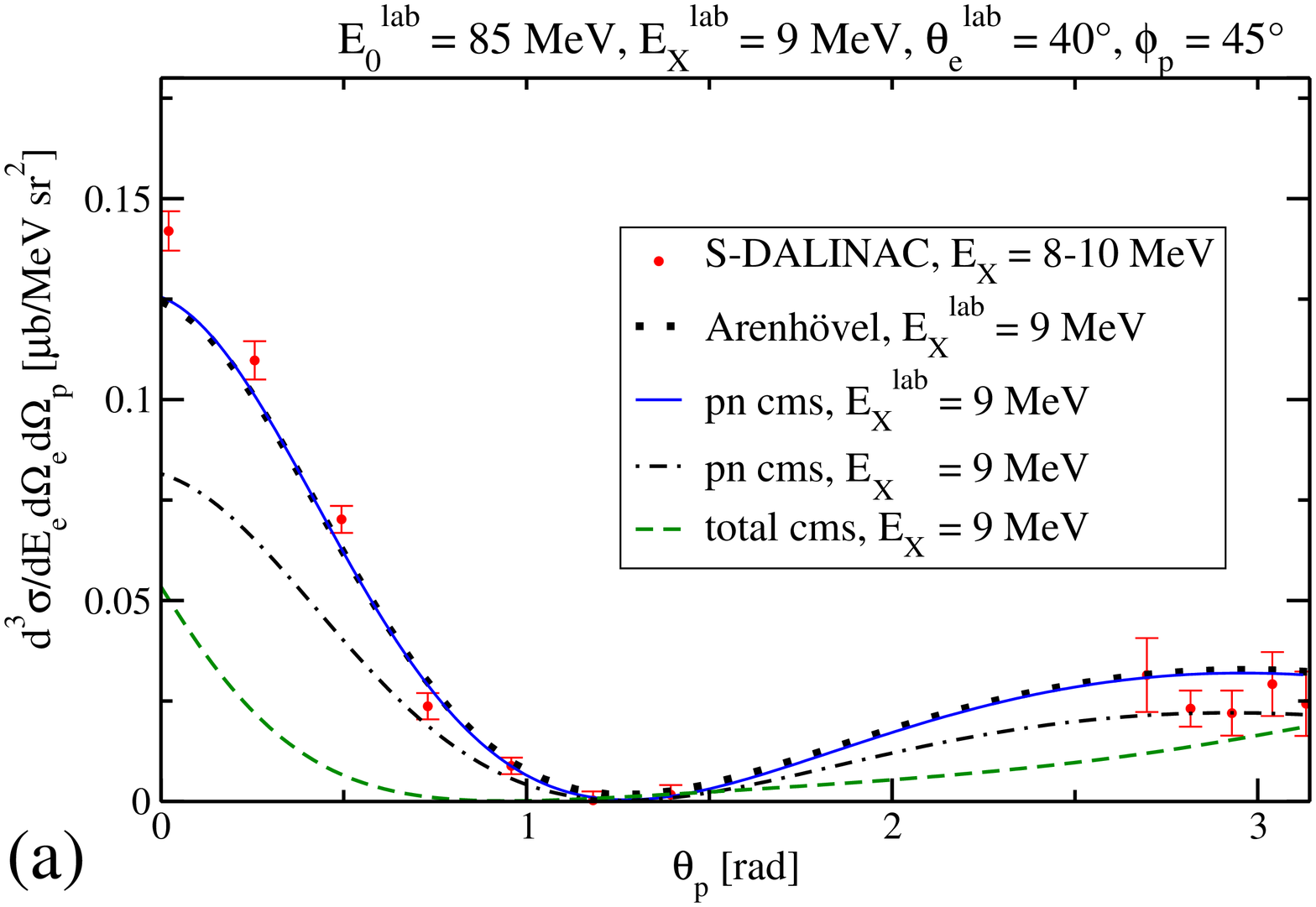}\\[2ex]
  \includegraphics[width=0.5\linewidth]{dsigma85764045ExVergl.eps}\\[2ex]
  \includegraphics[width=0.5\linewidth]{dsigma85764045Vergleichlog.eps}
  \caption{(Colour online) Examples of the triple-differential cross-section
    for $E_X^\mathrm{lab}=9\;\MeV$ at $E_0^\mathrm{lab}=85$~MeV, compared to
    \SDALINAC data~\cite{Exp} (with combined statistical and systematical
    error-bars) and to the result by Arenh\"ovel et
    al.~(squares)~\cite{ArenLeideTom04,Exp,Aren}. (a): Interpretation of
    ``$E_X=9\;\MeV$'' as given in the total cm frame, the $pn$-cm frame, or
    the lab-frame; (b): \EFTNoPion at $E_X^\mathrm{lab}=9$~MeV (solid line)
    and $E_X^\mathrm{(pn)-cm}=9$~MeV (i.e.~$E_X^\mathrm{lab}=10.6$~MeV,
    dashed).  The dotted curve is the double-differential cross-section per
    MeV, obtained by integrating over the bin
    $E_X^\mathrm{lab}\in[8;10]\;\MeV$; the difference to the solid line is
    almost invisible.  (c): comparison on a logarithmic scale of the
    \EFTNoPion contributions: electric transitions up to LO and \NXLO{2}
    respectively (recall that NLO is zero), and magnetic transitions up to
    NLO.}\label{cmcomp}
  \end{center}
\end{figure}

In Sect.~\ref{sec:kinematics}, we adopted the standard kinematics of
electro-disintegration according to which the leptonic part of the
cross-section is calculated in the lab frame (deuteron at rest), while the
hadronic part is determined in the proton-neutron cm-frame of the hadronic
sub-process $\gamma^* d\to pn$. In a slight but revealing abuse of language,
the latter is often referred to as ``the cm frame''.  According to a literal
reading of Ref.~\cite{Exp}, the \SDALINAC experiment would have been analysed
in the centre-of-mass frame of the \emph{whole} process, resulting in large
deviations between theory and experiment both in shape and normalisation even
of the full triple-differential cross-section, see Fig.~\ref{cmcomp}. We take
this to be a slip of the tongue and assume that the experiment was analysed
using standard kinematics~\cite{FabianAren,ArenLeideTom04,Arenhoevel:1992xu}.

Furthermore, the variable which labels the energy transferred by the photon in
the \SDALINAC experiment is said to be obtained ``after transformation to the
centre-of-mass system''~\cite{Exp}.  However, agreement of the data with both
the potential-model and \EFTNoPion calculation is found only if the
\SDALINAC-variable ``$E_X$'' refers to the lab frame. The point
$E_X^\mathrm{(pn)-cm}=9$~MeV would correspond to $E_X^\mathrm{lab}=10.6$~MeV
and a much smaller cross-section, so that the resulting curve would again
substantially deviate from data, see Fig.~\ref{cmcomp}.  We therefore
interpret ``$E_X$'' in \cite{Exp} as $E_X^\mathrm{lab}$, defined in the rest
frame of the deuteron target.

The experimental results are reported in 4 energy bins,
$E_X^\mathrm{lab}=[8;10]$, $[10;12]$, $[12;14]$ and $[14;16]\;\MeV$. We
interpret them such that the count-rate was normalised to give the
double-differential cross-section per MeV. Thus, we compare the
triple-differential cross-section for $E_X^\mathrm{lab}$ to the
double-differential one {\it per MeV}. The difference between averaging over
the $2\,\MeV$-range and taking the mean value of $E_X^\mathrm{lab}$ can be
neglected down to $E_X^\text{lab}=6\;\MeV$, as shown in the plot; see
also~\cite{diplom}. 
If $E_X$ is interpreted as given in the $pn$-cm frame \emph{and} the
cross-section as given over the whole $2\;\MeV$-bin, the data would
consistently over-shoot the theory results by $10$ to $20\%$.

We are grateful to H.~Arenh\"ovel~\cite{Aren} for confirming that the
potential-model results reported in~\cite{Exp} were determined in the same
kinematics used here.

\subsection{Total Differential Cross-Section}
\label{sec:datatotal}

\begin{figure}[!hptb]
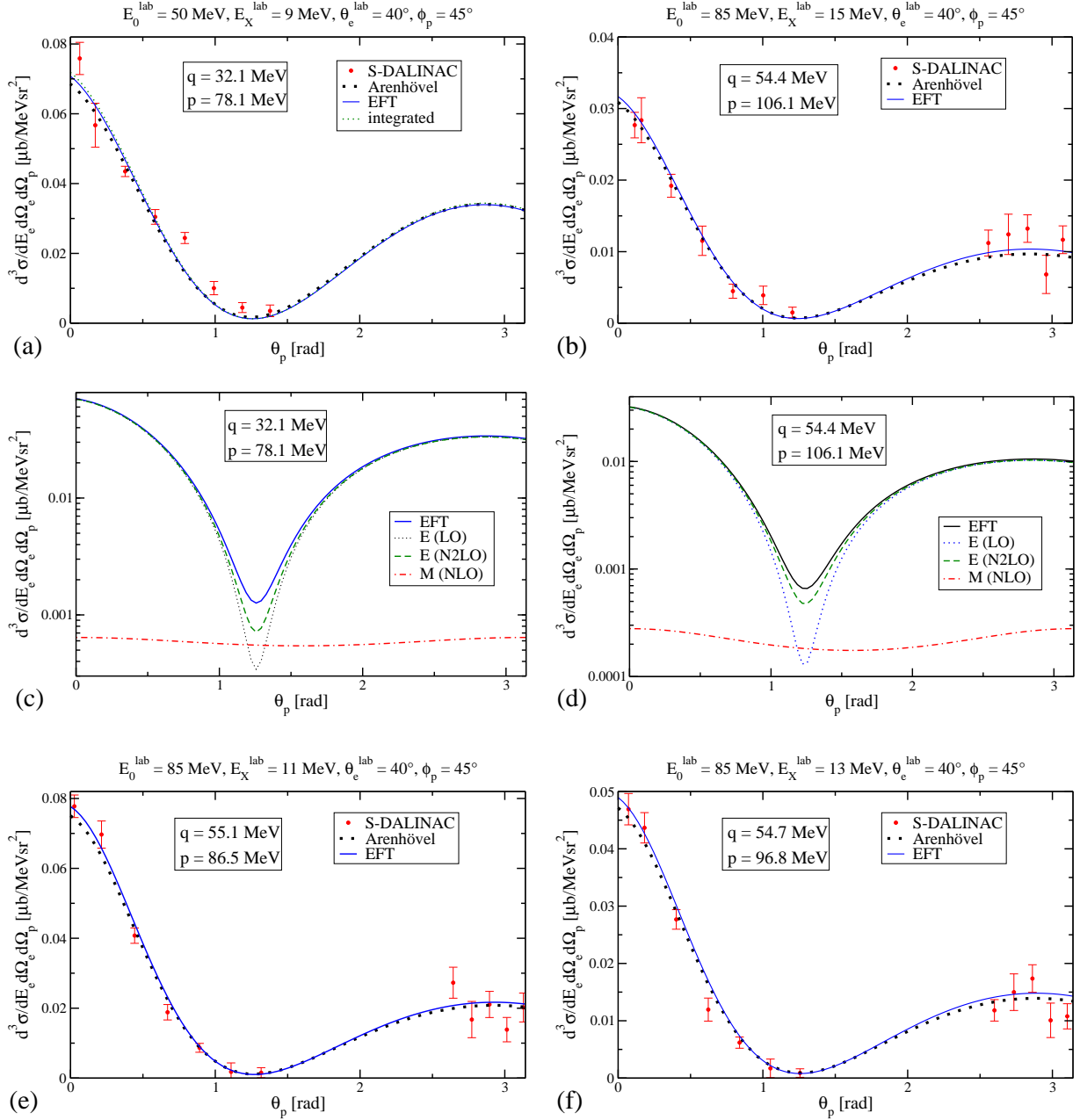

  \begin{center}
  \includegraphics[width=0.48\linewidth]{dsigma50414045.eps}\hq\hq
  \includegraphics[width=0.48\linewidth]{dsigma85704045.eps}\\[2ex]
  \includegraphics[width=0.48\linewidth]{dsigma50414045Vergleichlog.eps}\hq\hq
  \includegraphics[width=0.48\linewidth]{dsigma85704045Vergleichlog.eps}\\[3ex]
  \includegraphics[width=0.48\linewidth]{dsigma85744045.eps}\hq\hq
  \includegraphics[width=0.48\linewidth]{dsigma85724045.eps}
  \caption{(Colour online) \EFTNoPion results for the differential
    cross-section, compared to data and calculations by Arenh\"ovel et al.
    (a): data set at smallest energy and momentum transfer,
    $E_0^\text{lab}=50\;\MeV, \;E_X^\text{lab}=9\;\MeV$.  (b): data sets at
    largest energy and momentum transfer, $E_0^\text{lab}=85\;\MeV,
    \;E_X^\text{lab}=15\;\MeV$.  (e,f): data set at $E_0^\text{lab}=85\;\MeV,
    \;E_X^\text{lab}=\{11;13\}\;\MeV$. Comparisons of \EFTNoPion contributions
    are in Figs.~(c,d) included for the smallest and largest energy and momentum transfer:
    electric transitions up to LO and \NXLO{2} respectively (NLO is zero), and
    magnetic transitions up to NLO.}\label{dsigma}
  \end{center}
\end{figure}

Comparing in Fig.~\ref{dsigma} to the differential cross-sections reported in
Ref.~\cite{Exp} at $E_0^{\mathrm{lab}}=50$~MeV and $85$~MeV in several
$E_X^\mathrm{lab}$-bins confirms our interpretation of the experiment's
kinematical variables.  At $E_0^{\mathrm{lab}}=85$~MeV, additional data at
large $\Theta_p$ are available and within their (combined statistical and
systematical) error-bars compatible with both theoretical approaches. We note
excellent agreement with the calculation by Arenh\"ovel et al.~and with the
data within error-bars. In Fig.~\ref{dsigma}, a decomposition of the different
contributions in \EFTNoPion for the smallest and the largest energy and
momentum transfer of the \SDALINAC data is also shown.  As expected, the LO
electric transitions dominate except around the minimum at
$\Theta_p\approx70^\circ$, where NNLO contributions and magnetic transitions
play a significant r\^ole.  That means that by taking into account only
minimal coupling of photons to nucleon and di-baryon fields, disintegration
can be described highly accurately within \EFTNoPion. SD mixing, being the
only correction up to NNLO, contributes even less than estimated by power
counting.  This good convergence confirms our confidence that our results are
reliable. Less-known short-distance contributions play a very minor r\^ole,
see Sect.~\ref{sec:higherorder}.

The convergence plots also reveal an interesting facet: The \NXLO{2}
contribution has increased slightly for higher energy and momentum transfer,
but much less than expected from na\"ive dimensional analysis. Deviations
between the potential-model and \EFTNoPion results are also slightly more
pronounced there, especially at large $\Theta_p$. As typical momenta in the
process approach the breakdown-scale $\LambdaNoPion$ of \EFTNoPion, the
dimension-less parameter $Q=p_\text{typ}/\LambdaNoPion$ approaches unity and
the expansion is rendered useless. The momentum transfer $q\in[32;56]\;\MeV$
is comparable to the intrinsic low-momentum scale of the two-nucleon system,
namely the deuteron ``binding momentum'' $\gamma\approx45\;\MeV$. The momentum
of the final-state proton in the proton-neutron cm frame ranges however from
$75$ to $106\;\MeV$ and hence becomes at the upper end comparable to
$\LambdaNoPion\approx\mpi$. One would have expected this to affect in
particular the final-state interaction diagrams and the SD mixing
contributions at \NXLO{2}, which depend on $\vec{p}$ in the numerator.  The
observed enhancement is however much smaller.  The good convergence of
\EFTNoPion -- even for momenta close to the breakdown scale -- was found also
in many other applications, see e.g.~references
in~\cite{HadrtoNuc,Phillips,BedvKolck,hgritriton}.

\subsection{Comparing Structure Functions}
\label{sec:structure}

\begin{figure}[!htbp]
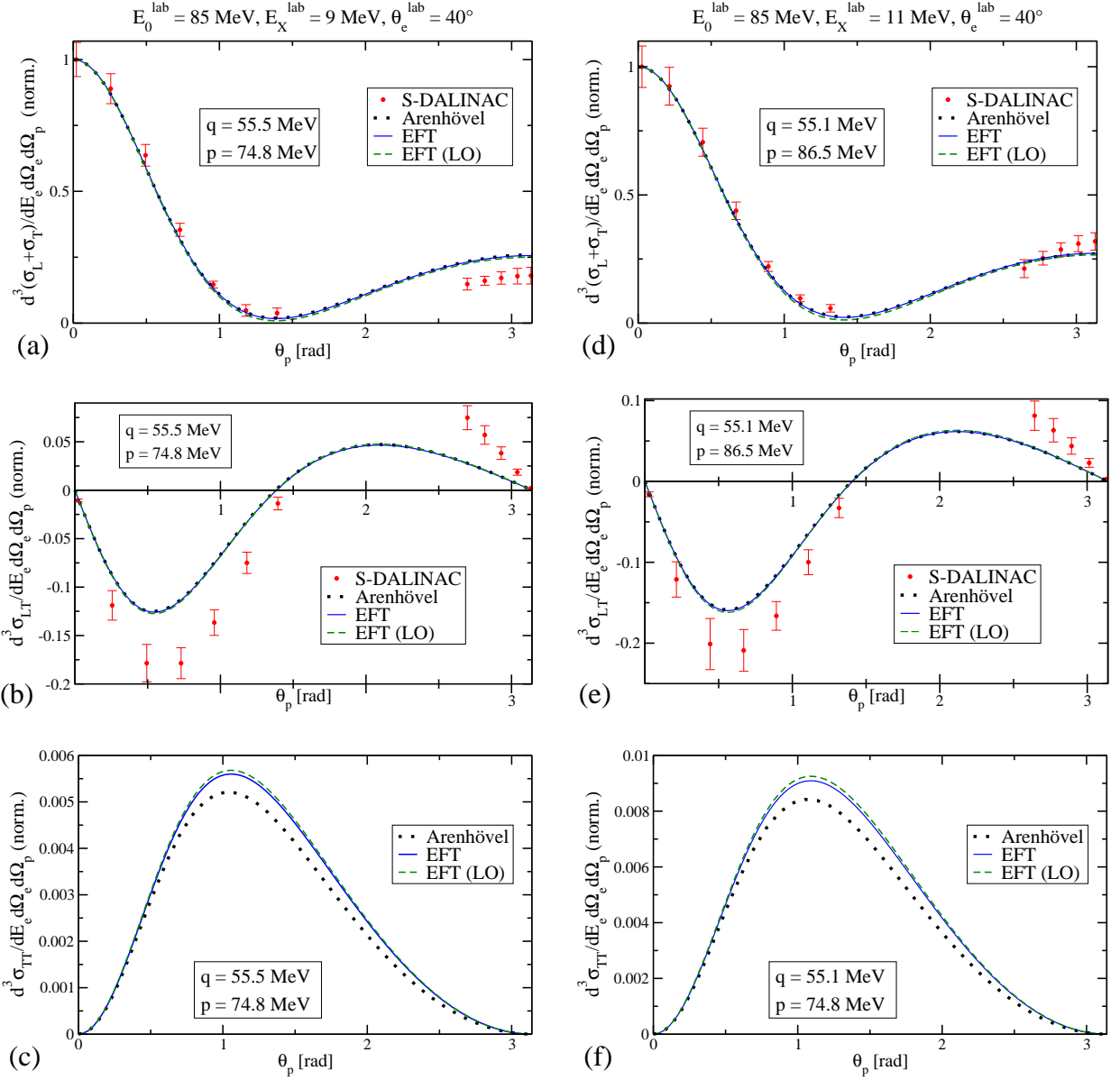

  \begin{center}
  \includegraphics[width=0.47\linewidth]{dsigmaLplusT857640norm.eps}\hq\hq\hq
  \includegraphics[width=0.47\linewidth]{dsigmaLplusT857440norm.eps}
\\[2ex]
  \raisebox{0.7ex}{\includegraphics[width=0.48\linewidth]{dsigmaLT857640norm.eps}}\hq\hq\hq
  \raisebox{0.7ex}{\includegraphics[width=0.48\linewidth]{dsigmaLT857440norm.eps}}
\\[2ex]
  \includegraphics[width=0.479\linewidth]{dsigmaTT857640norm.eps}\hq\hq\hq
  \includegraphics[width=0.479\linewidth]{dsigmaTT857440norm.eps}
%
  \caption{(Colour online) Decomposition of the triple-differential cross-section into (top to
    bottom) longitudinal-plus-transverse ($L+T$), longitudinal-transverse
    ($LT$) and transverse-transverse ($TT$) parts at
    $E_0^\mathrm{lab}=85$~MeV, normalised to $\sigma_{L+T}$ at $\Theta_p=0$.
    Left column: $E_X^{\mathrm{lab}}=9$~MeV; right column:
    $E_X^{\mathrm{lab}}=11$~MeV. No data are available for the $TT$
    interference cross-section.}\label{sigmadecomp}
  \end{center}
\end{figure}

\begin{figure}[!htbp]
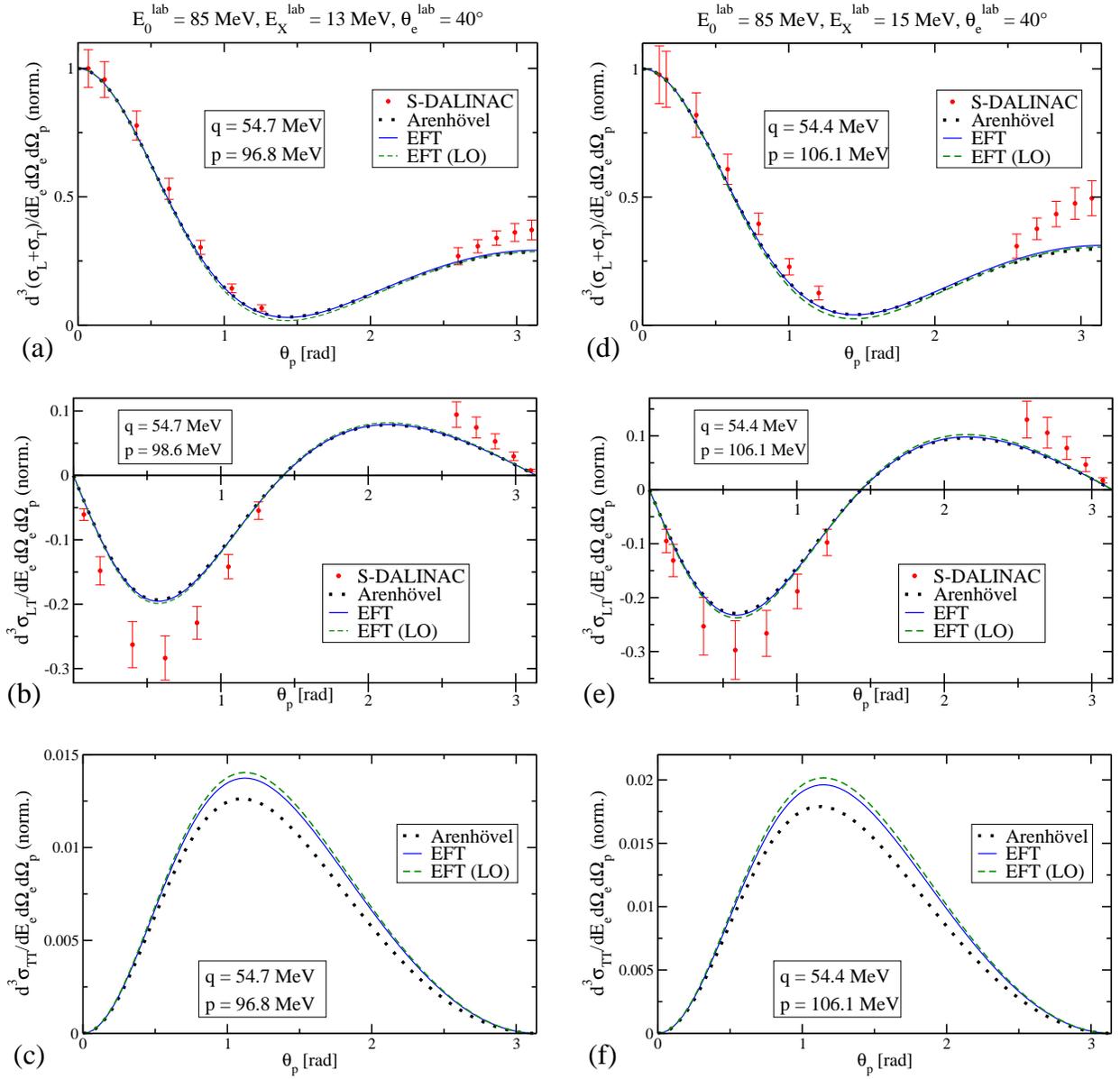

  \begin{center}
  \includegraphics[width=0.47\linewidth]{dsigmaLplusT857240norm.eps}\hq\hq\hq
  \includegraphics[width=0.47\linewidth]{dsigmaLplusT857040norm.eps}
\\[2ex]
  \raisebox{0.7ex}{\includegraphics[width=0.48\linewidth]{dsigmaLT857240norm.eps}}\hq\hq\hq
  \raisebox{0.7ex}{\includegraphics[width=0.48\linewidth]{dsigmaLT857040norm.eps}}
\\[2ex]
  \includegraphics[width=0.479\linewidth]{dsigmaTT857240norm.eps}\hq\hq\hq
  \includegraphics[width=0.479\linewidth]{dsigmaTT857040norm.eps}
%
  \caption{(Colour online) As Figure~\ref{sigmadecomp}, for $E_X^{\mathrm{lab}}=13$~MeV (left
    column); $E_X^{\mathrm{lab}}=15$~MeV (right column).\label{sigmadecomp2}}
  \end{center}
\end{figure}

The main goal of the \SDALINAC experiment~\cite{Exp} was a decomposition of
the contributions of different structure functions in \eqref{decomp}: the
longitudinal-plus-transverse cross-section, and the $LT$- and $TT$-interference
parts. However, the uncertainties in the data turned out to be too large for a
meaningful comparison of $\sigma_{TT}$ to theory. Studying the
interference terms provides information about the impact of final-state
interactions, since these can be $\Phi_p$-dependent, as re-iterated in
Sect.~\ref{sec:kinematics}. The contributions of the $L+T$, $LT$ and $TT$ terms are
shown in Figs.~\ref{sigmadecomp} and \ref{sigmadecomp2} for $E_0^{\mathrm{lab}}=85$~MeV and
$E_X^{\mathrm{lab}}=\{9;11;13;15\}\;\MeV$, again compared to data and the
potential-model result.

The results are normalised to $\dd^3(\sigma_L+\sigma_T)/(\dd
E_e^\mathrm{lab}\,\dd\Omega_e^\mathrm{lab} \,\dd\Omega_p)$ at
$\Theta_p=0^\circ$ to make the comparison independent of the absolute
normalisation of the data. This also mutes the question whether the
cross-sections were averaged or summed over each $E_X^\text{lab}$-bin. The
error of the value to which the data are normalised has been taken into
account in all error bars. $\sigma_L+\sigma_T$ dominates, whereas
$\sigma_{LT}$ and $\sigma_{TT}$ are one and two orders of magnitude smaller,
respectively. However, the interference terms become more relevant with
increasing energy transfer, reflecting the growing impact of final-state
interactions. In each figure, the result for the LO electric transitions in
\EFTNoPion is also given. Recall that in that case, $\sigma_{LT}$ and
$\sigma_{TT}$ are nonzero only because of the final-state interaction of the
proton with the virtual photon after the deuteron breakup,
Fig.~\ref{disLO}.(e1). The degree to which the $E1_V$-transition dominates --
already discussed in the previous Section -- is even more striking for
$\sigma_{LT}$. Magnetic transitions do not contribute at all to the
interference cross-sections, since their amplitudes are independent of
$\Phi_p$, see Sect.~\ref{sec:mag}.  The impact of $SD$ mixing is increasing
slightly with $E_X^{\mathrm{lab}}$, but is small for $\sigma_L+\sigma_T$ and
$\sigma_{TT}$, and almost negligible for $\sigma_{LT}$.

The \EFTNoPion results are again in good agreement with those of Arenh\"ovel
et al. Especially $\sigma_{LT}$ is essentially identical in both calculations.
Only in the very small quantity $\sigma_{TT}$ can a deviation be detected. It
is a factor of 4 larger than the difference between the LO and \NXLO{2}
results, which in turn is one way to estimate higher-order effects.  However,
we already hinted above that the contributions from SD mixing and magnetic
moment interactions are much smaller than estimated by na\"ive power-counting.
Another estimate for possible corrections is to set the uncertainty of the
\EFTNoPion calculation at \NXLO{2} as $\sim
Q^3=(p_\text{typ}/\LambdaNoPion)^3\approx [3\dots10]\%$ of the LO term,
assuming a typical momentum $p_\text{typ}\sim q\approx[45;60]\;\MeV$. This is
apt to over-estimate the theoretical uncertainties, as discussed in
Sect.~\ref{sec:higherorder} in conjunction with possible higher-order
corrections.  In short, all \EFTNoPion results should be understood with an
uncertainty which reflects an estimate of the higher-order effects not
considered.  A conservative accuracy limit would be $\lesssim10\%$ in each
observable.  It is thus safe to conclude that the potential-model and
\EFTNoPion results agree within the theoretical uncertainties.

\subsection{How To Resolve Discrepancies with Data?}
\label{sec:discrepancies}

\begin{figure}[!htb]
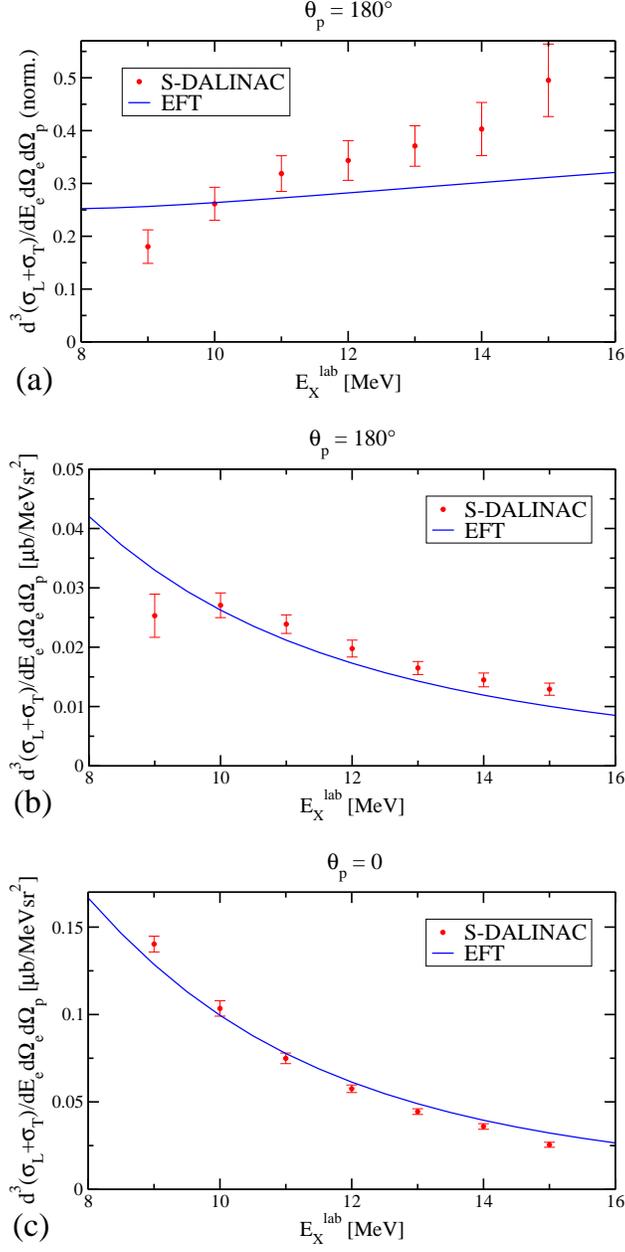

  \begin{center}
  \includegraphics[width=0.5\linewidth]{dsigmaLplusT85Ex180norm.eps}\\[2ex]
  \includegraphics[width=0.5\linewidth]{dsigmaLplusT85Ex180.eps}\\[2ex]
  \includegraphics[width=0.5\linewidth]{dsigmaLplusT85Ex00.eps}
  \caption{(Colour online) $E_X^{\mathrm{lab}}$-dependence of
    $\sigma_L+\sigma_T$ at $\Theta_p=180^\circ$, normalised to $\Theta_p=0^\circ$
    (a) and before normalisation (b). Bottom: $\sigma_L+\sigma_T$ at
    $\Theta_p=0^\circ$ (to which all data are normalised).  All graphs refer to
    $E_0^{\mathrm{lab}}=85$~MeV, $\Theta_e^{\mathrm{lab}}=40^\circ$.}\label{LplusTEx} 
  \end{center}
\end{figure}

The accordance of both theoretical approaches with the data, however, is
considerably worse.  The discrepancy ranges from just over one standard
deviation at the largest energy and momentum transfer, to three and more at
the smallest energy and momentum transfer. Not only does the calculated
longitudinal-transverse interference cross-section deviate by about $30 \%$
from the measured one (as pointed out in \cite{Exp}), but also
$\sigma_L+\sigma_T$ at large $\Theta_p$. In particular, it is remarkable that
the latter is over-predicted at $E_X^{\mathrm{lab}}=9$~MeV, but
under-predicted at $E_X^{\mathrm{lab}}=15$~MeV. A look at $\sigma_L+\sigma_T$
at $\Theta_p=180^\circ$ \emph{before data normalisation} reveals however a
reduced discrepancy, see Fig.~\ref{LplusTEx}. Only then do data and theory
agree on the shape of the curve, except at the smallest value of
$E_X^\text{lab}$, and an additional normalisation would solve the problem.
While the statistical significance of the discrepancy hardly changes, this may
indicate un-accounted systematic errors in the data.

Normalising to $\Theta_p=0$ seems to deteriorate the accord of calculations
and data -- although it avoids accounting for the absolute normalisation of
data. The reason can be seen by comparing
Fig.~\ref{LplusTEx}.(b) and (c): The cross-sections both at $\Theta_p=0$ and
$\Theta_p=180^\circ$ display relatively small deviations which go, however, in
different directions: The theoretical result e.~g.~at
$E_X^{\mathrm{lab}}=15$~MeV is a bit too small for $\Theta_p=180^\circ$, but
too large for $\Theta_p=0$. In dividing these two values, the discrepancy
increases.

\absatz The significant difference between data on the one hand, and the
agreeing theoretical predictions of \EFTNoPion and Arenh\"ovel et al.~on the
other hand, continues for the longitudinal-transverse interference cross
section $\sigma_{LT}$, see Figs.~\ref{sigmadecomp}.(b,e) and
\ref{sigmadecomp2}.(b,e). The angular dependence of the normalised values is
reproduced, but the normalisation differs between one standard deviations at
high energy and momentum transfer, and three standard deviations at high
energy and momentum transfer.

\begin{figure}[!htb]
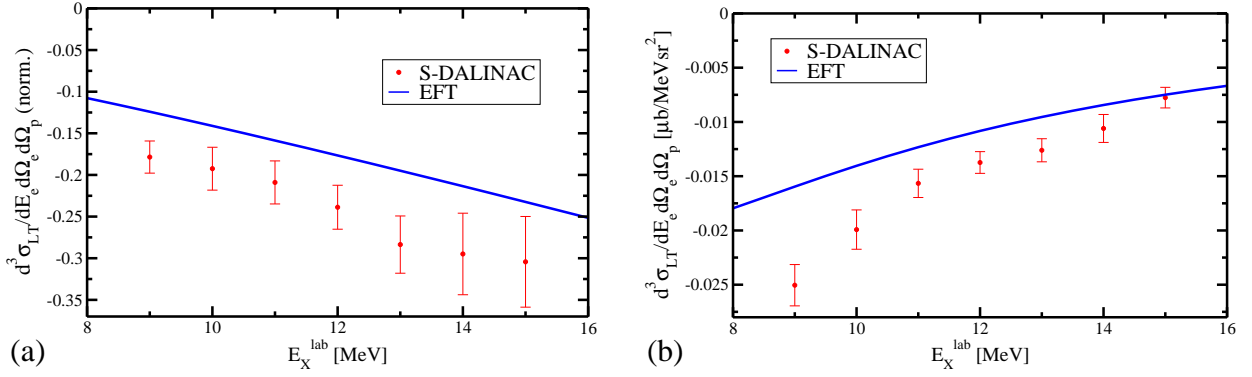

  \begin{center}
 \includegraphics[width=0.48\linewidth]{dsigmaLT85Ex35norm.eps}\hq\hq\hq
  \includegraphics[width=0.48\linewidth]{dsigmaLT85Ex35.eps}
  \caption{(Colour online) $E_X^{\mathrm{lab}}$-dependence of $\sigma_{LT}$ at the
    minimum around $\Theta_p=35^\circ$, normalised (left) and absolute values
    (right), both for $E_0^{\mathrm{lab}}=85$~MeV,
    $\Theta_e^{\mathrm{lab}}=40^\circ$.}
  \label{LTEx}
  \end{center}
\end{figure}

\begin{figure}[!htb]
  \begin{center}
 \includegraphics[width=0.35\linewidth,angle=-90]{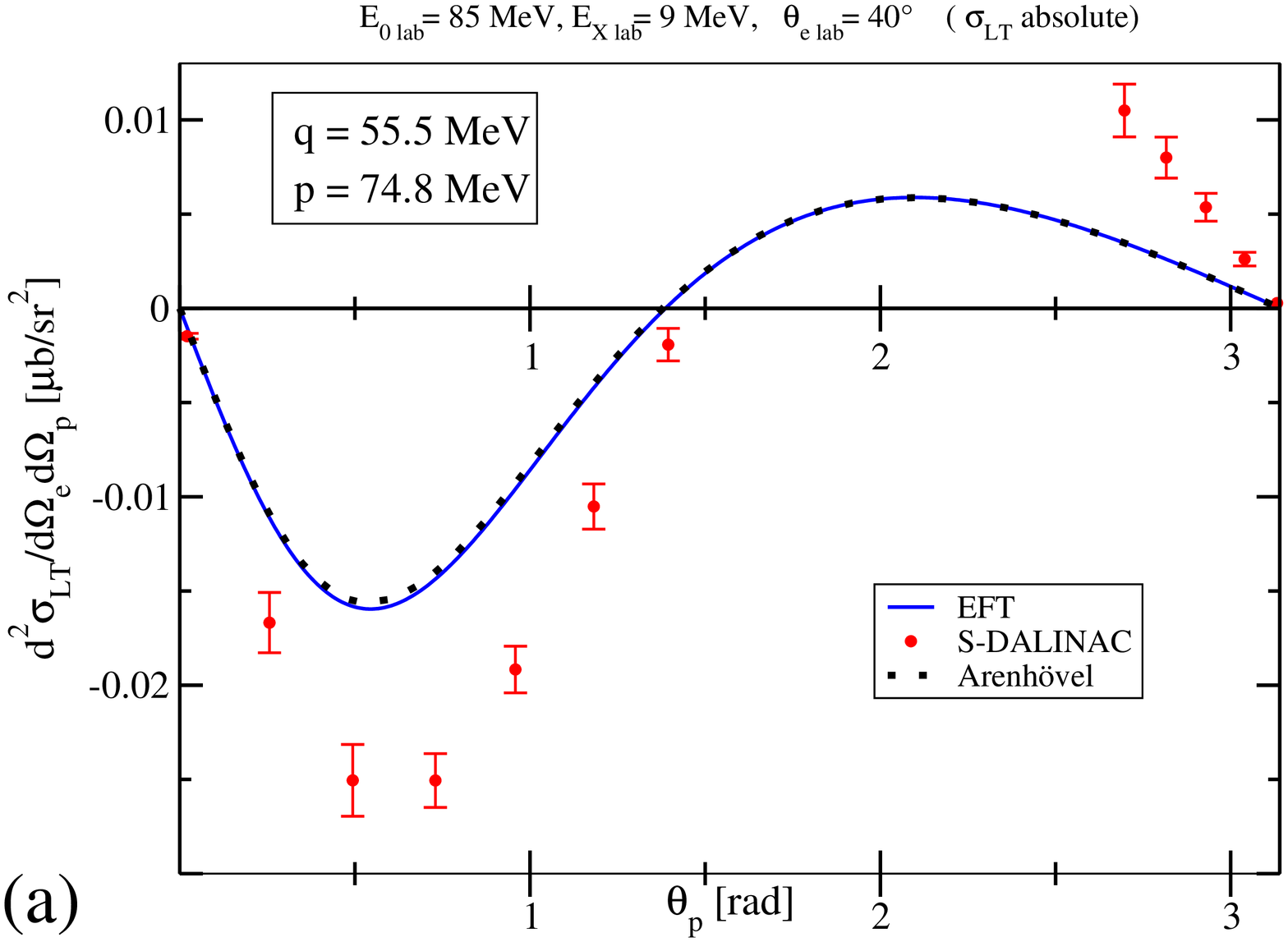}\hq\hq\hq
 \includegraphics[width=0.35\linewidth,angle=-90]{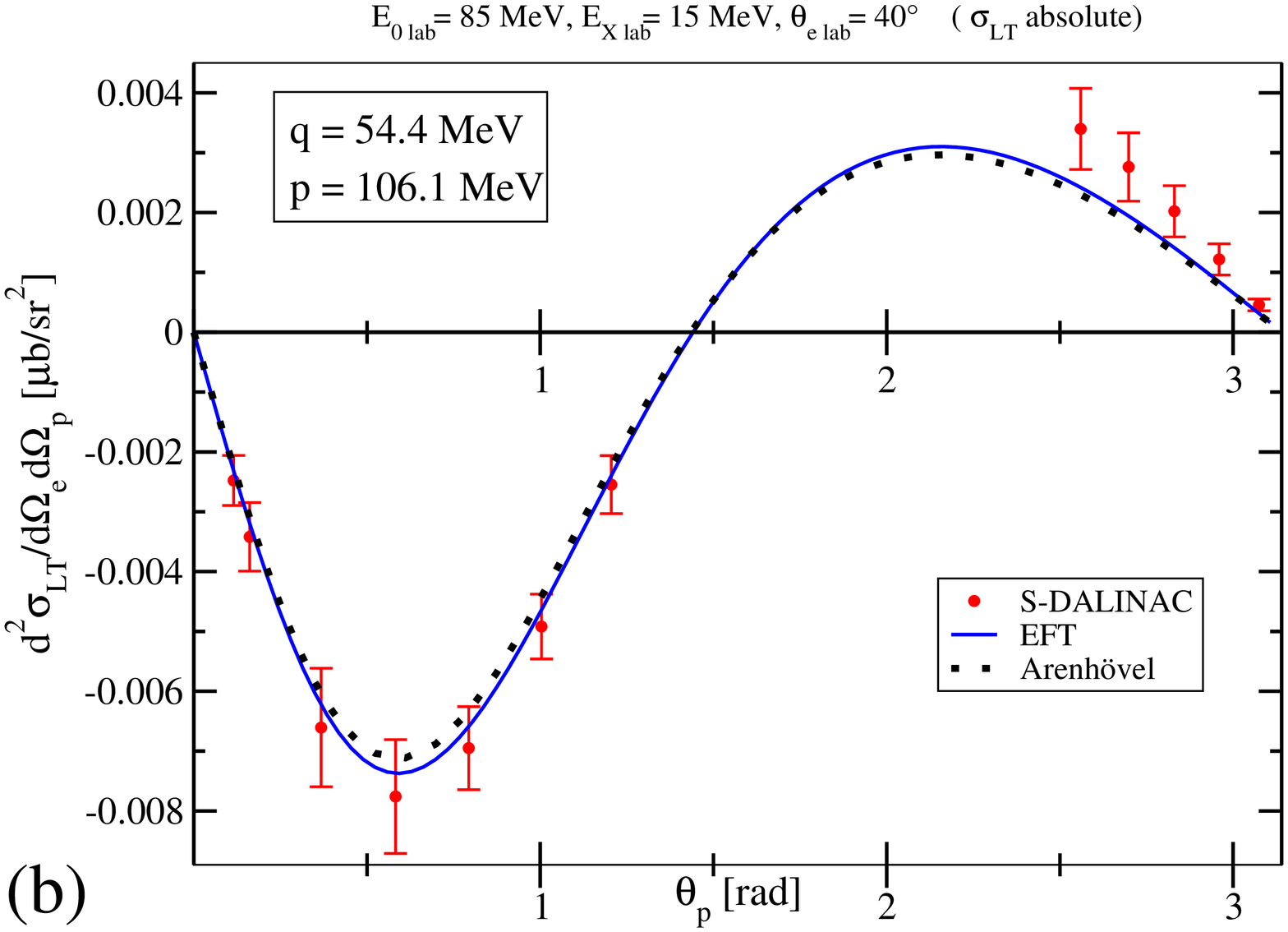}
  \caption{(Colour online) $E_X^{\mathrm{lab}}$-dependence of the \emph{absolute} values of
    $\sigma_{LT}$ for the extreme cases $E_X^{\mathrm{lab}}=9\;\MeV$ (left)
    and $15\;\MeV$ (right) at $E_0^{\mathrm{lab}}=85$~MeV,
    $\Theta_e^{\mathrm{lab}}=40^\circ$. Compare to the normalised data in
    Figs.~\ref{sigmadecomp} and \ref{sigmadecomp2}.}
  \label{LTEx2}
  \end{center}
\end{figure}

Figure~\ref{LTEx} examines the energy dependence of the discrepancy for
$\sigma_{LT}$ at the minimum around $\Theta_p=35^\circ$ as function of
$E_X^{\mathrm{lab}}$, both before and after normalisation of the data.
Again, the situation becomes, at least for higher energies, slightly better
when looking at the absolute values. However, a clear deviation remains, and
surprisingly becomes larger at smaller $E_X^{\mathrm{lab}}$. The shapes and
slopes of the un-normalised energy-dependence of the minima in $\sigma_{LT}$
do not match. The un-normalised data at high energy are, at least at
non-backward angles, compatible with the theoretical approaches within
error-bars, but the discrepancy at smaller energies and momentum transfers
worsens, see Fig.~\ref{LTEx2}.

\subsection{Discussion of Higher-Order Contributions}
\label{sec:higherorder}

In view of these discrepancies, one must investigate to which extend
additional interactions could remedy the problem. The \EFTNoPion result is
complete up to \NXLO{2} in electric and up to NLO in magnetic transitions,
resulting in a conservative accuracy-estimate of $\lesssim10\%$, as discussed
in Sect.~\ref{sec:structure}. Including some renormalisation-group invariant
sub-set of higher-order contributions provides a third alternative to estimate
the uncertainties of the calculation, besides order-by-order convergence and
\emph{prima facie} power counting.

However, the results so far provide a strong argument that higher-order
corrections cannot solve the problem: Recall first that $\sigma_{LT}$ is
nonzero only because of non-zero transverse components of the momentum
$\vec{p}$ of the outgoing proton. If, in spite of the power-counting, some
higher order contributions were important for $\sigma_{LT}$, they should
therefore contribute \emph{more} at higher energies, where $\vec{p}$ is bigger
and thus closer to the break-down scale $\LambdaNoPion$. This would lead to
larger higher-order corrections with increasing photon energy. This is however
at odds with our analysis of Figs.~\ref{sigmadecomp} to \ref{LTEx}. As
fine-tuning could circumvent this general argument, we explicitly consider in
the following some \NXLO{3} contributions.

Magnetic transitions are negligible at these energies and in these kinematics,
as seen in Figs.~\ref{cmcomp} and \ref{dsigma}.  Another contribution to
electric transitions includes $P$-wave nucleon-nucleon final state
interactions which also enter at \NXLO{3} \cite{ChenSavnpdg}. While they do
become stronger with increasing energy and momentum transfer, they are, even
at high energies, not even of the same order of magnitude as the \NXLO{2}
contributions.

Next, one might think that the correction to the deuteron quadrupole moment of
about $50\%$ provided by the $C_Q$-term in \eqref{Lsd} can solve the problem
at \NXLO{3}. However, this term is at best comparable in size to the \NXLO{2}
term from SD mixing. Its contribution will thus be even smaller than the
\NXLO{2} correction in Figs.~\ref{sigmadecomp} and \ref{sigmadecomp2}. An explicit
calculation~\cite{diplom} renders
\begin{equation}
  J^0_{ij} = -2y_tD_t\frac{C_Q}{M \rho_d} 
  \Big ({q}_i{q}_j-\frac{1}{3}\mathbf{q}^2 \delta_{ij} \Big)
  \;\;,\;\;\vec{J}_{ij}=0\;\;.
  \label{JCQ}
\end{equation}
The difference to the \NXLO{2} result is less than $0.5$~\% even at the
minimum around $\Theta_p=70^\circ$ at $E_0^\text{lab}=50\;\MeV,
\;E_X^\text{lab}=9\;\MeV$, where the effect would need to be biggest to remedy
the discrepancy. The contribution of this interaction to the interference
terms $\sigma_{LT}$ and $\sigma_{TT}$ is furthermore zero since \eqref{JCQ} is
independent of $\Phi_p$.

Corrections from relativistic kinematics are negligible. An example for a
relativistic effect which is dynamical but has been neglected in our
calculation is the spin-orbit interaction first considered in \EFTNoPion
by Chen et al.~\cite{ChenJiLi,Chen:2004wv},
\begin{equation}
  \mathcal{L}_{so} =\ii N^\dagger\left[\left(2\kappa_0-\frac{1}{2}\right) 
    +\left(2\kappa_1-\frac{1}{2}\right)\tau_3\right] \frac{e}{8M^2}
  \vec{\sigma}\cdot(\vec{D}\times\vec{E}-\vec{E}\times\vec{D})N\;\;,\label{so}
\end{equation}
where $\vec{E}$ is the electric field. This term is suppressed by
$p_\text{typ}/M\lesssim Q^2$ relative to the magnetic-field term in
$\mathcal{L}_N$ \eqref{LN} and thus enters in $M1_V$ transitions at \NXLO{4}.
Chean et al.~also demonstrated that it provides substantial angle-dependent
contributions to deuteron Compton scattering at energies around
$\omega\approx49\;\MeV$~\cite{Chen:2004wv}. Its Feynman diagrams are those of
the LO magnetic transitions, Fig.~\ref{Mdiag}.(m1-3), with the magnetic-moment
interaction substituted by \eqref{so}. The vector-component of the hadronic
current has the same structure as \eqref{curstrucM}, and the zero-component
can be written as
\begin{equation}
  J^{0\,(M1_V)}_\text{hadr} = e \sqrt{Z}\epsilon^{ijk}
  \epsilon_{(d)}^i\frac{1}{\sqrt{8}} (N_p^\dagger\sigma_2N_n^*) J^0_{jk}\;\;.
\end{equation}
The analytical results for these currents are reported in~\cite{diplom}. As
expected, they give only a small correction to the already negligible NLO
result for magnetic transitions. In $\sigma_{L+T}$, they lead to a correction
never exceeding $10^{-3}$ of the LO result. Although they are
$\Phi_p$-dependent, the relative difference to LO in $\sigma_{LT}$ is less
than $10^{-2}$, with $2\times10^{-3}$ at the extrema. In $\sigma_{TT}$, the
relative difference to LO at the maximum is $10^{-5}$ and never exceeds
$10^{-4}$.

We mention in passing also a \NXLO{4} contribution which can na\"ively be
incorporated by modifying the di-baryon propagators (\ref{dib3}/\ref{dib0})
to include the shape parameters $\rho_1=0.389\;\fm^3$ for the ${}^3S_1$
channel and $r_1=-0.48\;\fm^3$ for the ${}^1S_0$ channel of $NN$-scattering.
This neglects diagrams which come from gauging the derivative terms in the
Lagrangean which correspond to these corrections, so the result is not
complete.  However, we see that this leads only to minimal modifications of
the structure functions. $\sigma_{L+T}$ is increased by $\lesssim5\%$ at
back-angles, about $20\%$ in the minimum due to fine-tuning, and in general
much less. $\sigma_{LT}$ changes by less than $1\%$ at the extrema, and
$\sigma_{TT}$ only by $0.1\%$.

Overall, this discussion supports our estimate of the theoretical
uncertainty of our results.

\subsection{A Higher-Energy Experiment}
\label{sec:higherE}

\begin{figure}[t]
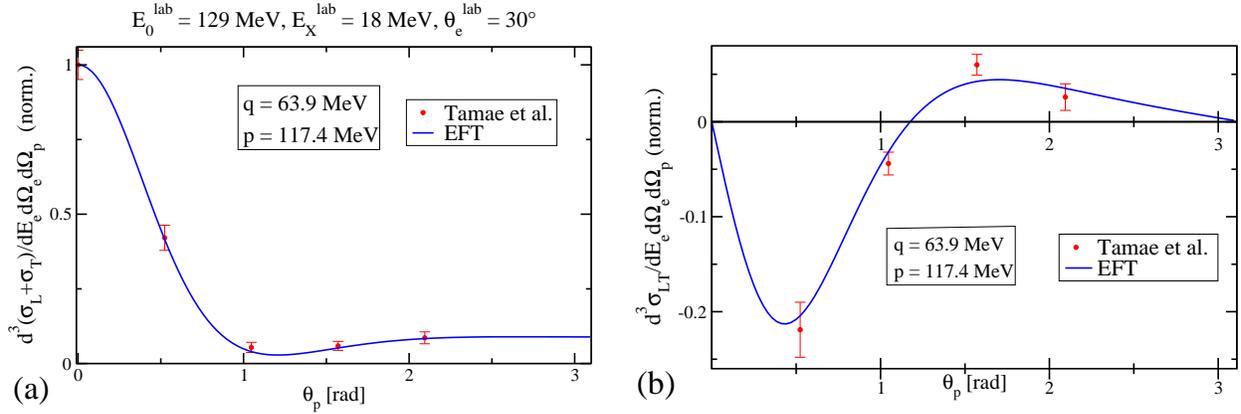

  \begin{center}
 \includegraphics[width=0.475\linewidth]{dsigmaLplusTlab12911130norm.eps}\hq\hq
  \raisebox{1.5mm}{
\includegraphics[width=0.49\linewidth]{dsigmaLTlab12911130norm.eps}}
\caption{(Colour online) Comparison to the data of Ref.~\cite{Tamae} for
  $\sigma_{L+T}$ (a) and $\sigma_{LT}$ (b), both in the {\it lab} frame (also
  for the hadronic variables!)  and normalised to $\sigma_{L+T}$ at
  $\Theta_p=0$.}\label{Tamae}
  \end{center}
\end{figure}

In Fig.~\ref{Tamae}, we finally compare the \EFTNoPion results to an earlier
experiment at a slightly higher momentum transfer \cite{Tamae} which reported
good accord between data and the potential model calculation by Arenh\"ovel et
al.\footnote{For this experiment, the calculations were performed using the
  Paris potential; the difference to the Bonn potential calculation in
  Ref.~\cite{ArenLeideTom04,Exp,Aren} should not be relevant here.}~also for
the normalised $\sigma_{LT}$.
Although the involved proton momentum $p=117$~MeV is even closer to the
breakdown scale of \EFTNoPion than for the \SDALINAC kinematics, we observe
very good agreement for both $\sigma_{L+T}$ and $\sigma_{LT}$.  It is
remarkable that the result describes the data so well in a r\'egime in which
\EFTNoPion may start to become unreliable, but deviates significantly from
data at lower momenta.

\section{Concluding Questions}
\setcounter{equation}{0}
\label{sec:conclusions}

In a parameter-free \NXLO{2} calculation using \EFTNoPion, a manifestly gauge
and renormalisation-group invariant, well-tested formulation of few-nucleon
physics with analytical results which are rooted in a systematic,
model-independent low-energy expansion of all nuclear forces, with an \emph{a
  priori} estimate of theoretical uncertainties corroborated by order-by-order
convergence, the \SDALINAC data on deuteron electro-disintegration
$d(e,e^\prime p)n$ at low energy and momentum transfer~\cite{Exp,Expweb}
cannot be explained.

Since an important aspect of any Effective Field Theory is its universality and
resulting model-independence, this leads us to a conclusion beyond the
\EFTNoPion approach: The predictions of models which reproduce or share the
input of \EFTNoPion must agree with its results to within the accuracy of the
EFT calculation, in the range where \EFTNoPion is applicable.  The analysis in
Sect.~\ref{sec:data} showed that the \SDALINAC data at low momentum and energy
transfer lie undoubtedly within this r\'egime.  Therefore, any self-consistent
potential model, irrespective of the detail led treatment of meson-exchange
currents, ``off-shell effects'', cutoff-dependence, etc.~will to the accuracy
outlined here agree with the \EFTNoPion result, if it incorporates the same
ingredients: the same deuteron binding energy, the same asymptotic
normalisation $Z$ and asymptotic S-to-D wave ratio $\eta_{sd}$ of the deuteron
wave-function, the same scattering length and effective range in the $^1S_0$
channel, the same iso-vectorial magnetic moment of the nucleon $\kappa_1$ and
the same total radiative thermal capture cross-section of neutrons on protons.
This is nicely confirmed for electro-disintegration at low energies: The
potential-model approach by Arenh\"ovel et al.~\cite{ArenLeideTom04,Exp,Aren}
yields essentially the same results as \EFTNoPion, especially for
$\sigma_{LT}$.

We find furthermore that $\sigma_{L+T}$, $\sigma_{LT}$ and $\sigma_{TT}$ are
all dominated up to the few-percent level by the leading-order electric
transition, i.e.~by minimally coupling the photon to the nucleon and deuteron.
This coupling is sensitive only to the asymptotics of the deuteron
wave-function. The statement is strongest for $\sigma_{LT}$, which shows the
discrepancy: It is dominated up to the 1\%-level by only one LO electric
process, Fig.~\ref{disLO}.(e1), i.e.~by the asymptotic normalisation $Z$ of
the deuteron wave-function, eq.~\eqref{deutwavefu}. In contradistinction, the
discrepancy to the \SDALINAC data amounts to up to 3 standard-deviations or
$30\%$.

These findings suggest a re-analysis of the experiment~\cite{ShevCosel}.  We
identified some questions concerning the kinematics and systematics of the
experimental analysis and caution that the differences could in part arise
from a dis-advantageous normalisation of the data. If the discrepancies were
confirmed, this would pose a highly non-trivial problem for Nuclear Theory.
Only new data for deuteron electro-disintegration near threshold and at low
momentum transfers can settle the issue definitively, particularly concerning
the decomposition into the contributions of different structure functions. We
are however confident to maintain that the validity and error-estimate of the
$E1_V$-part of the photo-dissociation cross-section relevant for Big-Bang
Nucleo-Synthesis as calculated in
\EFTNoPion~\cite{ChenRupSavnpdg,ChenSavnpdg,Ando:2005cz,Rupaknpdg} are not in
question.


\section*{Acknowledgements}
We am indebted to D.~B.~Kaplan for pointing out to us the puzzle of the
\SDALINAC data when they were published. P.~von Neumann-Cosel provided
excellent insight into the experiment. Both he and H.~Arenh\"ovel worked
closely with us to resolve the reference-frame issue, answering a barrage of
questions with great patience. H.~Arenh\"ovel provided invaluable clues by his
diligence and by the potential-model results he kindly provided at our
request.  M.~Schwamb helped with comments and further observations, and
W.~Parke with a careful reading of this manuscript. Our thanks also to
N.~Kaiser and W.~Weise for critical companionship and a careful reading of the
manuscript of the Diplom thesis~\cite{diplom} on which this article is based.
H.W.G.~gratefully acknowledges the Institut f\"ur Theoretische Physik (T39) at
TU M\"unchen and \SDALINAC at TU Darmstadt for its kind hospitality.
This work was supported in part by the Bundesministerium f\"ur Forschung und
Technologie, by the Deutsche Forschungsgemeinschaft under contracts GR1887/2-2
and 3-1, by the CAREER-grant PHY-0645498 of the National Science Foundation
and by a Department of Energy grant DE-FG02-95ER-40907.

\newpage

\appendix
\section{Currents and Integrals}
\setcounter{equation}{0}
\label{app:appendix}

We now list the analytical results for the hadronic currents described in
Sect.~\ref{sec:el} and \ref{sec:mag}.

\subsection{Useful Loop Integrals}
\label{app:integrals}

The following loop integrals are calculated by the methodology outlined
in~\cite{SavSpr,FlemMehStew}\footnote{We use a slightly different notation.},
using contour integration for the energy part and the PDS subtraction scheme
combined with dimensional integration in $D$ spatial
dimension~\cite{KapSavWisePDS,KapSavWisePC} to identify divergences,
parameterised by $\mu$, and tensorial reduction. The reader may
consult~\cite[App.~D]{diplom} for details.

The fundamental bubble-sum integral is~\cite{KapSavWisePDS,KapSavWisePC}
\begin{equation}
  I_0^{(1)}(a)=\left(\frac{\mu}{2}\right)^{3-D} \int \deintdim{D}{l} 
  \frac{1}{(\vec{l}+\frac{\vec{q}}{2})^2+a}\stackrel{\mbox{\scriptsize{PDS}}}{=}
  \frac{1}{4\pi}(\mu-\sqrt{a}) \label{I01}\;\;.
\end{equation}
Reference~\cite{SavSpr} provides
\begin{equation}
  I_0^{(2)}(a,b;q)=\int \frac{\dd^3l}{(2\pi)^3} \frac{1}{l^2+a}
  \frac{1}{(\vec{l}+\frac{\vec{q}}{2})^2+b} =
  \frac{1}{2\pi q} \arctan
  \left(\frac{q}{2(\sqrt{a}+\sqrt{b})}\right)\;\;.  \label{I02}
\end{equation}
From that, one finds the $\mu$-independent result with one loop
momentum in the numerator
\begin{equation}
  A_1(a,b;q)=\int \frac{\dd^D}{(2\pi)^D} \frac{\vec{l}\cdot\vec{q}}{l^2+a}
	       \frac{1}{(\vec{l}+\frac{\vec{q}}{2})^2+b}=
               I_0^{(1)}(a)-I_0^{(1)}(b)-\Big(\frac{q^2}{4}+b-a\Big)
	       I_0^{(2)}(a,b;q)
\end{equation}
and the following $\mu$-dependent ones with two powers of loop momenta in the
numerator:
\begin{eqnarray}
  A_2(a,b;q) &=&\frac{1}{D-1}\int \deintdim{D}{l}
      \frac{l^2-\frac{(\vec{l}\cdot\vec{q})^2}{q^2}}{l^2+a}
      \frac{1}{(\vec{l}+\frac{\vec{q}}{2})^2+b}\nonumber \\
  &= &\frac{1}{1-D}\frac{1}{q^2} 
      \left[\frac{2-D}{2}I_0^{(1)}(b)-aI_0^{(2)}(a,b;q)
      +D\frac{\frac{q^2}{4}+b-a}{q^2} A_1(a,b;q) \right]
    \\
   B_2(a,b;q) &=&\frac{1}{1-D}\frac{1}{q^2} \int \deintdim{D}{l}
      \frac{l^2-D\frac{(\vec{l}\cdot\vec{q})^2}{l^2}}{l^2+a}
      \frac{1}{(\vec{l}+\frac{\vec{q}}{2})^2+b} \nonumber \\
      &=&\frac{1}{1-D}\frac{1}{q^2} 
      \left[\frac{2-D}{2}I_0^{(1)}(b)-aI_0^{(2)}(a,b;q)
      +D\frac{\frac{q^2}{4}+b-a}{q^2} A_1(a,b;q) \right]
\end{eqnarray}
Three powers of loop momenta  in the numerator are covered by the integrals
\begin{eqnarray}
  A_3(a,b;q)
  &= &\frac{1}{(D-1)q^2}\int \deintdim{D}{l} 
  \frac{1}{l^2+a} \frac{1}{(\vec{l}+\frac{\vec{q}}{2})^2+b}
  \left[(\vec{q}\cdot\vec{l})l^2-\frac{1}{q^2}(\vec{q}\cdot\vec{l})^3
  \right] \nonumber \\
  &= &\frac{1}{(D-1)q^2} \left[-\frac{q^2}{2}I_0^{(1)}(b) -aA_1(a,b;q)
              -\frac{1}{q^2} \hat{A}_3(a,b;q)\right]\;\;, \label{A3}\\[1.5ex]
  B_3(a,b;q)&= &\frac{D+2}{D-1}\frac{1}{q^6}\int \deintdim{D}{l} 
  \frac{}{l^2+a} \frac{1}{(\vec{l}+\frac{\vec{q}}{2})^2+b}
  \left[(\vec{q}\cdot\vec{l})^3-\frac{3q^2}{D+2}(\vec{q}\cdot\vec{l})
  l^2\right] \nonumber \\
  &= &\frac{D+2}{D-1}\frac{1}{q^6} \left[\hat{A}_3(a,b;q)-\frac{3q^2}{D+2}
  \left(-\frac{q^2}{2}I_0^{(1)}(b)-aA_1(a,b;q)\right)\right]\;\;, \label{B3}
\end{eqnarray}
with the aid of
\begin{eqnarray}
  \lefteqn{\hat{A}_3(a,b;q) =\int \deintdim{D}{l} 
  \frac{(\vec{q}\cdot\vec{l})^3}{l^2+a} 
  \frac{1}{(\vec{l}+\frac{\vec{q}}{2})^2+b} } \nonumber\\
  &=&-\frac{aq^2}{D}I_0^{(1)}(a)- 
  \left(\frac{q^4}{4}-\frac{bq^2}{D}\right)I_0^{(1)}(b) \nonumber \\
  & &
  +\left(\frac{q^2}{4}+b-  a\right) 
  \left[-\frac{q^2}{2}I_0^{(1)}(b)+\left(\frac{q^2}{4}+b- a\right)
    A_1(a,b;q)\right]\;\;.
\end{eqnarray}

\subsection{Electric Currents up to \NXLO{2}}
\label{app:electric}

The electric contributions to the hadronic current at LO are labelled as in
Fig.~\ref{disLO}. The 4-vector contribution is for the first diagram  with
$\vec{r}:= 2\vec{p}-\vec{q}$
\begin{eqnarray}
  J_{ij}^{0\,\text{(e1)}} = -2y_tD_p(\vec{p}; \omega,\vec{q})\;\delta_{ij}&;&
  \vec{J}_{ij}^{\text{(e1)}} =J_{ij}^{0\,\text{(e1)}} \frac{\vec{r}}{2M}\;\;,
\end{eqnarray}
where the proton propagator 
\begin{equation}
  \ii D_p(\vec{p}; \omega,\vec{q}) = \frac{\ii}{-\omega
    +\frac{\vec{p}\cdot\vec{q}}{M}-\frac{q^2}{2M}}\;\;; 
       \label{protprop}
\end{equation}
could be approximated for real photons by dropping the $q^2$-term in the
denominator ($\omega=q$), but not in electro-disintegration because the
kinematics imposes $\omega\sim q^2/M$.

The second diagram gives
\begin{eqnarray}
  J_{ij}^{0\,\text{(e2)}} =2y_tD_t(p)\;\delta_{ij} &;&
  \vec{J}_{ij}^{\text{(e2)}} =-J_{ij}^{0\,\text{(e2)}} \frac{\vec{q}}{4M}\;\;,
\end{eqnarray}
with the spin-triplet di-baryon propagator 
\begin{equation}
  \ii D_t(p) = \frac{M\rho_d}{2}
  \frac{\ii}{\gamma-\frac{\rho_d}{2}(\gamma^2+p^2)+\ii p}\;\;.\label{dib3}
\end{equation}
After integrating over the loop, diagram (3) results in
\begin{eqnarray}
  J_{ij}^{0\,\text{(e3)}} 
  &= &-2y_t^3D_t(p)M^2I_0^{(2)}(-p^2,\gamma^2)\;\delta_{ij}\;\;,  \label{J03} \\
  \vec{J}_{ij}^{\text{(e3)}}&=& y_t^3
  D_t(p) M\left[I_0^{(2)}(-p^2,\gamma^2)+\frac{2}{q^2}A_1(-p^2,\gamma^2;q)\right] 
  \vec{q}\;\delta_{ij}\;\;. \label{Jv3}
\end{eqnarray}
Here, as in the following, $-p^2$ is understood to be $-p^2-\ii\epsilon$ with
$\epsilon\searrow0$, i.~e.~$\sqrt{-p^2}=-\ii p$, where $p$ is the proton momentum.

\absatz The \NXLO{2} contributions come from $\mathrm{SD}$-mixing operators
only. Following the numeration of Fig.~\ref{sddiag}, one finds:
\begin{eqnarray}
  J_{ij}^{0\text{(sd1)}} =-\frac{1}{2}\frac{C_{sd}}{\sqrt{M\rho_d}}
  D_p(\vec{p}; \omega,\vec{q})
  \Big({r}_i{r}_j-\frac{1}{3}\vec{r}^2\delta_{ij}\Big) &,&
  \vec{J}_{ij}^{\text{(sd1)}} =J_{ij}^{0(sd1)}\frac{\vec{r}}{2M}\;\;,
\end{eqnarray}
\begin{eqnarray}
  J_{ij}^{0\text{(sd2)}} =\frac{C_{sd}}{\sqrt{M\rho_d}}D_t(p)
  \Big({p}_i{p}_j-\frac{1}{3}\vec{p}^2\delta_{ij}\Big)&,&
  \vec{J}_{ij}^{\text{(sd2)}} =-J_{ij}^{0(sd2)}\frac{\vec{q}}{4M}\;\;,
\end{eqnarray}
\begin{equation}
  J_{ij}^{0\text{(sd3)}}
  =-2\frac{C_{sd}y_t^2}{\sqrt{M\rho_d}}D_t(p)M^2I_0^{(2)}(-p^2,\gamma^2;q) 
  \Big({p}_i{p}_j-\frac{1}{3}\vec{p}^2\delta_{ij}\Big)
\end{equation}
\begin{equation}
  \vec{J}_{ij}^{\text{(sd3)}} =\frac{C_{sd}y_t^2}{\sqrt{M\rho_d}} D_t(p) 
  M\left[I_0^{(2)}(-p^2,\gamma^2;q)+\frac{2}{q^2}A_1(-p^2,\gamma^2;q)\right]
  \Big({p}_i{p}_j-\frac{1}{3}\vec{p}^2\delta_{ij}\Big)\vec{q}
\end{equation}
\begin{equation}
  J_{ij}^{0\text{(sd4+sd5)}} =-2\frac{C_{sd}y_t^2}{\sqrt{M\rho_d}}D_t(p) M^2
  \left[B_2(-p^2,\gamma^2;q)+B_2(\gamma^2,-p^2;q)\right]
  \Big({q}_i{q}_j-\frac{1}{3}\vec{q}^2\delta_{ij}\Big)\;\;, 
\end{equation}
\begin{eqnarray}
  \left(\vec{J}_{ij}^{\text{(sd4+sd5)}}\right)_k
  &=&2\frac{C_{sd}}{\sqrt{M\rho_d}}D_t(p)y_t^2M
  \left[\frac{1}{2} B_2(-p^2,\gamma^2;q)
    \Big({q}_i{q}_j-\frac{1}{3}\vec{q}^2\delta_{ij}\Big)
    {q}_k\right. \nonumber \\
  &&+[A_3(-p^2,\gamma^2;q)-A_3(\gamma^2,-p^2;q)]\left({q}_i\delta_{jk}
    +{q}_j\delta_{ik}-\frac{2}{3}{q}_k\delta_{ij}\right)\nonumber\\
  &&\left.+[B_3(-p^2,\gamma^2;q)-B_3(\gamma^2,-p^2;q)]
    \left({q}_i{q}_j{q}_k- 
      \frac{1}{3}q^2{q}_k\delta_{ij}\right)\right]\;\;.
\end{eqnarray}
The last three diagrams contribute only to the three-vector component of the
current: 
\begin{equation}
  \left(\vec{J}_{ij}^{\text{(sd6)}}\right)_k =\frac{C_{sd}}{\sqrt{M\rho_d}}
  \left[-{p}_i\delta_{jk}+
  \Big(-{p}_j+\frac{1}{2}{q}_j\Big)\delta_{ik}
  +\frac{2}{3}\Big({p}_k-\frac{1}{4}{q}_k\Big)\delta_{ij}\right]
\end{equation}
\begin{equation}
  \left(\vec{J}_{ij}^{\text{(sd7+sd8)}}\right)_k =
  \frac{C_{sd}}{\sqrt{M\rho_d}}D_t(p) y_t^2\frac{M}{4}
  [I_0^{(1)}(\gamma^2)-I_0^{(1)}(-p^2)] 
  \left({q}_j\delta_{ik}+{q}_i\delta_{jk}
  -\frac{2}{3}{q}_k\delta_{ij}\right)
\end{equation}
With these results, it is simple to show that the LO and \NXLO{2} electric
currents are separately gauge-invariant, $q_\mu J^\mu_{ij}=0$. It is also
noteworthy that the total current at each order is independent of the
regularisation parameter $\mu$.

\subsection{Magnetic Currents up to NLO}
\label{app:magnetic}

The LO diagrams, numbered as in Fig.~\ref{Mdiag}, contribute
\begin{equation}
  \vec{J}^\text{(m1)} =-y_t D_p(\vec{p};\omega,\vec{q}) \frac{\kappa_1}{M}\;\vec{q}
  \;\;,\;\;
  \vec{J}^\text{(m2)} =-y_t D_n(\vec{p};\omega,\vec{q}) \frac{\kappa_1}{M}\;\vec{q}
\end{equation}
\begin{equation}
  \vec{J}^\text{(m3)} =-2 y_t D_s(p) y_s^2 \kappa_1 M
  I_0^{(2)}(-p^2,\gamma^2;q)\;\vec{q}\;\;.
\end{equation}
with neutron propagator $ D_n(\vec{p};\omega,\vec{q})=
D_p(-\vec{p};\omega,\vec{q})$ and spin-singlet di-baryon propagator 
\begin{equation}
  \ii D_s(p) = \frac{Mr_0}{2} \frac{\ii}{\frac{1}{a_0}-\frac{r_0}{2}p^2+\ii
    p}\;\;.\label{dib0}
\end{equation}
The current provided by the NLO contribution is 
\begin{equation} 
  \vec{J}^\text{(m4)} = -2 y_s D_s(p) \frac{L_1}{M\sqrt{r_0\rho_d}}\; \vec{q}\;\;.
\end{equation}
Notice that these currents are all transversal only. 

\newpage

\end{document}